\documentclass[prd,aps,nofootinbib,preprint,eqsecnum]{revtex4}

\usepackage{hhline}
\usepackage{latexsym,graphicx}
\usepackage{subfigure}
\usepackage{tikz}
\usepackage[percent]{overpic}
\usepackage{pgfplots}
\pgfplotsset{compat=1.18}
\usepackage{amssymb}
\usepackage{amsmath}
\usepackage{amscd}
\usepackage{amsthm}
\usepackage{float}
\usepackage{xcolor}
\usepackage{comment}

\begin{document}
\title{Quintessence-dominated cyclic universe with negative cosmological constant}

\author{Nasr Ahmed$^1$}
\email[]{nasr.ahmed@nriag.sci.eg}
\author{Kazuharu Bamba$^2$ }
\email[]{bamba@sss.fukushima-u.ac.jp}
\affiliation{$^1$ Astronomy Department, National Research Institute of Astronomy and Geophysics, Egypt.\\ $^2$ Faculty of Symbiotic Systems Science, Fukushima University, Fukushima 960-1296, Japan.}
\begin{abstract}
We investigate two simplified non-singular cyclic models with a negative time-varying cosmological constant to represent the non-conventional mechanism of negative cosmological constant expected to address the late-time cosmic acceleration. We show that a physically acceptable evolution with positive energy density can be realized, while negative energy density dominates in case of a positive or zero cosmological constant. In the first model, we demonstrate a sign flipping of the cosmic pressure in a quintessence-dominated universe with no violation of the null energy condition. In the second model, we propose a matter-bounce scenario with showing the crossing of the phantom divide line in the vicinity of the bounce. We find that while we get positive kinetic term and scalar potential, the sum of scalar and quantum potentials is negative. 
\end{abstract}
\keywords{Modified gravity, Dark energy, Cyclic cosmology}
\maketitle

\section{Introduction}

\quad In cyclic cosmology, the universe has an infinite sequence of cycles of expansion and contraction where there is no beginning or end of time. In searching for solutions for the problems of the standard cosmology such as flatness and initial singularity, cyclic scenarios (also named as bouncing cosmology) have been firstly proposed as a serious alternative to the cosmic inflation theory. In such singularity-free bouncing models, which have earned an extensive interest in the literature \cite{bounc2,bounc22,bounc33,bounc3, bounc5}, the contraction-expansion cycle continues forever.\par

\quad Cyclic models have been investigated in modified gravity \cite{bounc5,bounc7b,bounc9,bounc10,bounce101}, a general model in $f(R)$ gravity has been introduced in Ref.~\cite{33a}. The Conformal Cyclic Cosmology (CCC) \cite{Penroseccc} assumes a cosmic succession of expanding phases named as cosmic aeons. Each aeon emerges with its own big bang and ultimately participates in an indefinitely expanding de Sitter-like state. A unitary version of CCC has been suggested in Ref.~\cite{33g}. A two-branes cyclic Model, inspired by M-theory, has been presented in Ref.~\cite{58} in which the equation of state parameter satisfies $w \gg 1$ during the phase of contraction. The two-branes cyclic model (also called the ekpyrotic model \cite{ekpyro1} or bouncing branes theory) proposes that a big bang-like event is produced by the periodic collision of two parallel, three-dimensional branes (universes) in a higher-dimensional space. The immense kinetic energy from the collision is converted into the matter, radiation, and heat that characterize the initial conditions of the new universe's hot and dense phase. With an exponential growth of the scale factor from one cycle to another, a new cyclic theory has been suggested in Ref.~\cite{new} that leads to a nearly scale-invariant density perturbations' spectrum. \par

\quad The Anti-de Sitter (AdS)-de Sitter (dS) transition is corresponding to a sign switching cosmological constant \cite{cosmc}. The possibility of existence of a negative cosmological constant (i.e., AdS vacua or AdS ground state) in the dark energy sector has been recently discussed in Ref.~\cite{negativelambda1}. In view of the measurement of  the current value of the Hubble parameter $H_0$ from the Hubble Space Telescope HST and SH0ES (Supernovae and $H_0$ for the Equation of State of dark energy) team, They showed that this negative $\Lambda$ (which naturally arises in string theory) can be consistent with cosmological observations. A quintessence model with negative $\Lambda$ has been constructed and compared with the $\Lambda$CDM model through a different combination of Cosmic Microwave Background (CMB) Radiation, SnIa, Baryon Acoustic Osculation (BAO) \cite{DESI:2025zpo,DESI:2025zgx} and $H_0$ data. Such comparison showed that quintessence models with a negative $\Lambda$ is either preferred over the $\Lambda$CDM or performs equally as it. \par

\quad The presence of a dS vacuum presents a problem for the standard cosmology where it implies a universe that eventually becomes an empty and eternally accelerating. While our late-time universe appears to be approaching a dS vacuum (which implies a positive $\Lambda$), constructing a stable dS vacua within string theory is extremely difficult. Many attempts have been performed based on uplifting mechanisms but their consistency is debated. The most well-known uplifting scenarios are KKLT and LVS, where the stability of the dS vacuum has been extensively investigated in the last twenty years \cite{nasrian}. In Ref.~\cite{negativelambda2} it has been demonstrated that all dS vacua obtained from uplifting AdS stable vacua have associated unsolvable Galois groups. Consequently, all dS vacua lack analytic solutions. 
Recently, the role of AdS vacua in the evolution of our universe has been explored, particularly in connection with inflation in the AdS landscape \cite{role3,role4} and the Hubble tension, which could be related the AdS vacuum \cite{role1, role2} (see Ref.~\cite{role5} and the references therein). \par
\quad Although negative $\Lambda$ can't drive cosmic acceleration, it can exist at the same time with a positive dark energy component. A survey of the corresponding consequences has been conducted in light of CMB, pre-DESI BAO dataset \cite{negativelambda1,role8}, and recent James gebb Space Telescope (JWST) observations \cite{role9,role10}. The first search for a negative $\Lambda$, based on recent DESI BAO measurements \cite{desi11, desi12} combined with Planck CMB and Pantheon Plus supernova data, was performed in Ref.~\cite{role5}, where indications of a negative $\Lambda$ were reported. Considering a universe with negative vacuum energy, a novel mechanism for explaining inflation through cyclic phases was proposed in Ref.~\cite{role11}.\par

\quad The AdS vacuum corresponds to a negative cosmological constant, the natural prediction of negative $\Lambda$ by string theory \cite{stringc} contradicts the need for positive $\Lambda$ based on observations. There is an early attempt to solve this discrepancy in the framework of string theory but leads to unstable worlds. Another attempt is the $\Lambda_s$CDM model cosmology which is a modified $\Lambda$CDM cosmology with a sign flipping cosmological constant $\Lambda_s$ \cite{role13}, i.e., $\Lambda \rightarrow \Lambda_s=\Lambda_{s0} sgn[z_{+}-z]$. where $\Lambda_{s0}>0$ and $z_+$ is the redshift at which $\Lambda$ changes sign. However, the sign switching of the cosmological constant in $\Lambda_s$ cosmology is introduced manually. Other attempts to achieve the AdS–dS transition include quintessence fields with a negative $\Lambda$ \cite{negativelambda1}.\par

\quad The current work represents an attempt to probe cosmic evolution with negative time-varying cosmological constant in the context of conformal Bohm-de Broglie gravity. The paper is organized as follows. In Sec. \ref{Scale-invariance gravity}, we review scale-invariant gravity and its cosmological implications. In Sec. \ref{secprho}, we study a cyclic solution of the cosmological equations in flat space. The evolution of the potential and kinetic terms is analyzed in Sec. \ref{seckv}. Energy conditions are examined in Sec. \ref{conditions}. In Sec. \ref{bounce1}, we present a second bouncing model. Conclusions are given in Sec. \ref{conclusion}.
\section{Scale-invariance gravity} \label{Scale-invariance gravity}
\quad Scale-invariance implies that the equations remain unchanged under the transformation $ds^{'}=\lambda(t) ds$ of the line element, with $ds^{'}$ is the GR line element and the line element $ds$ belongs to a more general space \cite{basic01}. Even though several modified scale-invariance gravity theories have been proposed to account for cosmic acceleration \cite{basic0,basic1, basic}, the ability of scale-invariance cosmology to fully describe cosmic evolution remains unverified. Based on the Bohmian trajectory approach, an alternative interpretation of quantum mechanics has been used in Ref.~\cite{basic} to show that it is possible to integrate quantum effects into the classical equations of motion by applying a conformal transformation to the background metric. The conformal transformation is given as
\begin{equation} \label{conf}
\widetilde{g}_{\mu\nu}=e^Q g_{\mu\nu} ~, ~ Q=\frac{3}{2} \left(q-\frac{1}{2}\right)\frac{H^2}{M^2},
\end{equation}
where $Q$ is the quantum potential, $q=\frac{\ddot{a}a}{\dot{a}^2}$ is the deceleration parameter, $H=\frac{\dot{a}}{a}$ is the Hubble parameter and $M$ is the particle's mass. The modified Einstein equations are \cite{basic0,basic1}
\begin{equation} \label{ein22}
\widetilde{R}_{\mu\nu}-\frac{1}{2}\widetilde{g}_{\mu\nu}\widetilde{R}=8\pi G \widetilde{T}_{\mu\nu}+\Lambda \widetilde{g}_{\mu\nu},
\end{equation}
where $\widetilde{R}_{\mu\nu}$ is the Ricci tensor with respect to the modified metric $\widetilde{g}_{\mu\nu}$, $\widetilde{R}$ is the Ricci scalar, $\Lambda$ is the effective cosmological constant, and $\widetilde{T}_{\mu\nu}= \widetilde{T}^{(\mathrm{M})}_{\mu\nu}+ \widetilde{T}^{(\mathrm{Q})}_{\mu\nu}$ is the energy-momentum tensor where $\widetilde{T}^{(\mathrm{M})}_{\mu\nu}$ is related to the matter contribution, and $\widetilde{T}^{(\mathrm{Q})}_{\mu\nu}$ originates from the quantum potential's energy density. The Friedmann‐Lemaître‐Robertson-Walker (FLRW) metric metric is given by
\begin{equation}
ds^{2}=-dt^{2}+a^{2}(t)\left( \frac{dr^{2}}{1-\kappa r^2}+r^2d\theta^2+r^2\sin^2\theta d\phi^2 \right) \label{RW}
\end{equation} 
where $r$, $\theta$, $\phi$ are comoving coordinates, $a(t)$ is the scale factor, $t$ is time, $\kappa$ is either $0$, $-1$ or $+1$ for flat, open and closed universe respectively. Applying the transformation Eq.~(\ref{conf}) to the Robertson-Walker metric results in the following modified equations for a Universe made up of scalar, spin-zero, massive particles.
\begin{eqnarray}
\frac{\dot{a}^2}{a^2}&=&\frac{8\pi G}{3}\rho+\frac{\Lambda \lambda^2}{3}-2\frac{\dot{\lambda}\dot{a}}{\lambda a}+\frac{\dot{\lambda}^2}{\lambda^2}-\frac{\kappa}{a^2}, \label{RW1} \\    
\frac{\ddot{a}}{a}&=&-\frac{4\pi G}{3}(\rho+3p)+\frac{\Lambda \lambda^2}{3}-\frac{\dot{\lambda}\dot{a}}{\lambda a}-\frac{\dot{\lambda}^2}{\lambda^2}-\frac{\ddot{\lambda}}{\lambda}. \label{RW2}
\end{eqnarray}
where $\lambda^2=e^Q$, the ordinary FLRW equations can be obtained for $\lambda=1$. As has been pointed out in Ref.~\cite{basic}, the extra terms in these modified equations appear as an acceleration in cosmic expansion. In this work, we consider only the observationally supported flat case \cite{flatt,nas222}. While in Ref.~\cite{basic} a universe with zero cosmological constant has been considered, we consider a more general case of a time-dependent cosmological constant $\Lambda(t)$. Some decay laws of $\Lambda$ have been suggested in Ref.~\cite{decaylaws} as 
\begin{equation} 
\Lambda = A t^{-l}, ~~~ \Lambda = B a^{-m}, ~~~ \Lambda = C H^{n}, ~~~ \Lambda = D q^{r}.
\end{equation}
where $A$, $B$, $C$, $D$, $l$, $m$, $n$ and $r$ are adjustable constants, $a$ is the scale factor, and $q$ is the deceleration parameter. other models have been suggested for a varying cosmological constant in terms of the Hubble parameter \cite{20d,20d2}
\begin{eqnarray} 
\Lambda(H)&=& \lambda_{0} +\alpha_{0} H + 3 \beta_{0} H^2, \label{vary2} \\
\Lambda(H)&=&  3 \alpha H^2 +\beta a^{-2},
\end{eqnarray} 
where $\lambda_{0}$, $\alpha_{0}$, $\beta_{0}$, $\alpha$ and $\beta$ are constants. The linear choice $\Lambda(H)=\sigma H$ where $\sigma > 0$ has been suggested earlier in Refs.~\cite{lin11,lin12} to help explaining cosmic acceleration. However, the analysis in Refs.~\cite{lin13,lin14} showed that this linear choice is inconsistent with observational data.\par
\quad The generalized cosmological constant model in Eq.~(\ref{vary2}) was initially proposed in Ref.~\cite{20d} and subsequently examined in detail in Refs.~\cite{lin15,lin16,lin17} using observational data. The authors showed that models with $\lambda=0$ encounter significant problems, as they fail to agree with both the observational constraints and the linear growth rate of structure formation. In contrast, models with $\lambda \neq 0$ remain compatible with the data and approach a $\Lambda$CDM–like behavior at late times.

\section{Negative $\Lambda(H)$ Solution} \label{secprho}

\quad Numerous physical arguments in the literature support adopting a model-independent approach to construct cosmological models and study dark energy dynamics \cite{adopt1,adopt2,adopp3,adopp4}. In this section, we follow the same parametrization approach for cosmological parameters to solve the cosmological equations. Although the majority of parametrizations in the literature concentrate on the deceleration parameter and the equation of state parameter \cite{adopp}, other physical or geometrical quantities can also be parametrized. A 4-dimensional FLRW cyclic cosmology with quintom matter has been constructed in Ref.~\cite{basicc} making use of the following sign-changing periodic form of the deceleration parameter in compatible with the observationally suggested cosmic transit
\begin{equation} \label{q}
q(t)=-\frac{\ddot{a}a}{\dot{a}^2}=m \cos kt -1, 
\end{equation}
where $m$ and $k$ are positive constants. This form leads to the following scale factor which avoids the future big rip (BR) singularity  
\begin{equation} \label{a}
a(t)=A \exp \left[ \frac{2}{\sqrt{c^2-m^2}}\arctan \left(\frac{c\tan \left(\frac{kt}{2}\right)+m}{\sqrt{c^2-m^2}}\right)\right],
\end{equation}
where $A$ and $c$ are integration constants. Singularities such as big rip \cite{futurefate} can be avoided for a specific choice of $c>m$ \cite{basicc}. The scale factor represented in Eq.~(\ref{a}) implies a new type of periodic behavior in which the Hubble parameter $H$ oscillates but remains positive. Since the cyclic solution represented by Eq.~(\ref{a}) is non-singular, it allows for the unification of the late-time cosmic acceleration and inflation. Hence, in the present work, We have adopted an empirical approach to solve the cosmological equations based on a specific observationally-restricted form of $q(t)$. \par

\quad By making a slight modification of Eq.~(\ref{vary2}), we suggest a new formula for a negative time-varying cosmological constant given as    
\begin{equation} \label{vary3}
\Lambda(H)= -(\lambda_{0} +\alpha_{0} H + 3 \beta_{0} H^2).
\end{equation}
As we have mentioned earlier, there exists a large class of models with a generic time-dependent cosmological constant, commonly referred to as $\Lambda(t)$-cosmologies. However, as the time dependence of $\Lambda(t)$ is typically introduced through ad hoc parametrization, many of these models are purely phenomenological. Running vacuum models are theoretically motivated models where vacuum energy density evolves according to the renormalization group in quantum field theory in curved space-time. So, Instead of an arbitrary $\Lambda(t)$, the vacuum energy is expressed as a function of the Hubble parameter $\rho_{\Lambda}(H)=c_0+\nu H^2 + \alpha H^4$ where the dependence on $H$ comes from quantum corrections. Since $H=H(t)$, all running vacuum models are time-dependent vacuum models. While in the current work we assume a time-varying cosmological constant, this assumption is not derived from the action but is adopted as a phenomenological description of a dynamical vacuum energy which is a widely used approach in cosmology \cite{rvm1,rvm2}.\par

\quad The proposed negative $\Lambda(H)$ ansatz (\ref{vary3}) is motivated by both theoretical and observational considerations that support the possibility of a negative cosmological constant. First, frameworks such as string theory naturally admit solutions with negative $\Lambda$ \cite{stringc}. Second, negative $\Lambda$ can halt eternal acceleration in cyclic cosmology as it offers a significant attractive effect at large scales, facilitating turnaround and re-collapse. In contrast, positive $\Lambda$ generically leads to eternal acceleration and inhibits cyclic behavior \cite{dd4}. Third, the ansatz is also supported by insights from the AdS/CFT correspondence \cite{ah}, which provides a consistent theoretical setting involving negative vacuum energy. \par

\quad While the standard $\Lambda$CDM model assumes a constant positive vacuum energy ($\Lambda > 0$), there has been a significant surge in theoretical and observational interest in a negative cosmological constant (AdS) over the last few years. The possibility of constraining negative $\Lambda$ using the post-reionization HI 21-cm power spectrum has been studied in Ref.~\cite{bb} through investigating the quintessence models with the most common DE equation of state parameterization and adding a non-vanishing vacua. It has been found that, by BOSS (SDSS) data analysis, it appears possible to incorporate a negative $\Lambda$ into a phantom dark energy model. As has been pointed out in Ref.~\cite{role5} that negative $\Lambda$ can still coexist with one evolving component of positive dark energy, although cosmic acceleration cannot be driven by it. the related impacts have been examined in light of cosmic microwave background, pre-DESI BAO dataset and recent JWST observations.\par

\quad The non-conventional mechanism of negative cosmological constant that are anticipated to address cosmic acceleration has been discussed by many authors and in different contexts \cite{dd4, gardena, grand, gu, land, ma, ba}. The role of negative $\Lambda$ in solving the problem of eternal acceleration has been discussed in Ref.~\cite{dd4}. After contrasting the model against the Hubble diagram of Type Ia supernovae, it has been found that it can fit quite well a large dataset which means that a negative $\Lambda$ is indeed allowed. A strong argument for the negative $\Lambda$ has been given by the AdS/CFT correspondence \cite{ah}. A stable de Sitter solution with negative $\Lambda$ has been found in Ref.~\cite{ma} by investigating gravitational theories with the Gauss-Bonnet squared term. The role of negative $\Lambda$ in the context of black hole physics both in AdS and dS background has been probed in Ref.~\cite{BHch}. Hence, the negative time-dependent cosmological constant in the present work has a strong base in the literature. \par

\begin{figure}[t]
  \centering         
	\subfigure[$\Lambda$]{\label{F00003}\includegraphics[width=0.3\textwidth]{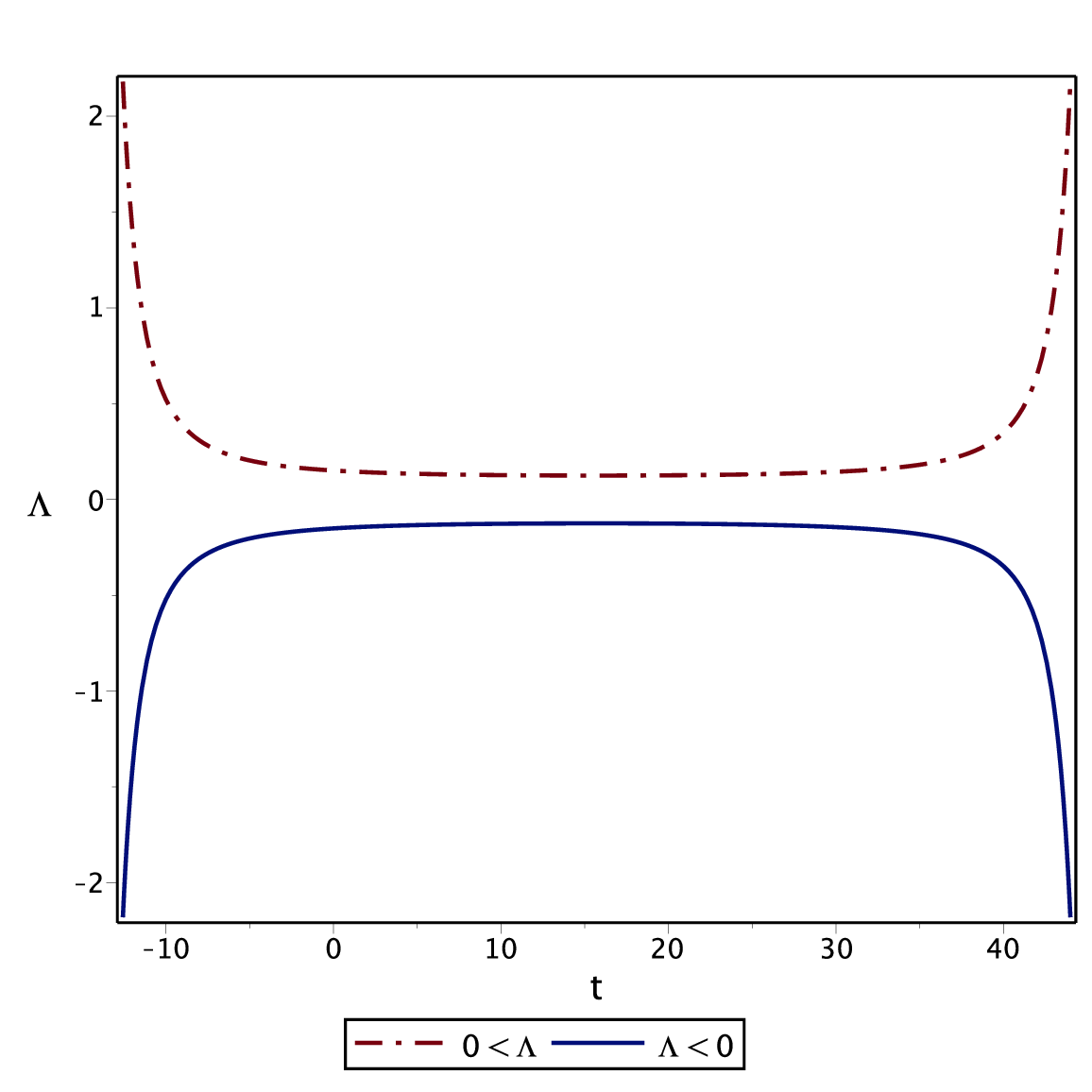}}
 \hspace{0.5cm}
	  \subfigure[$\rho$]{\label{F63}\includegraphics[width=0.3\textwidth]{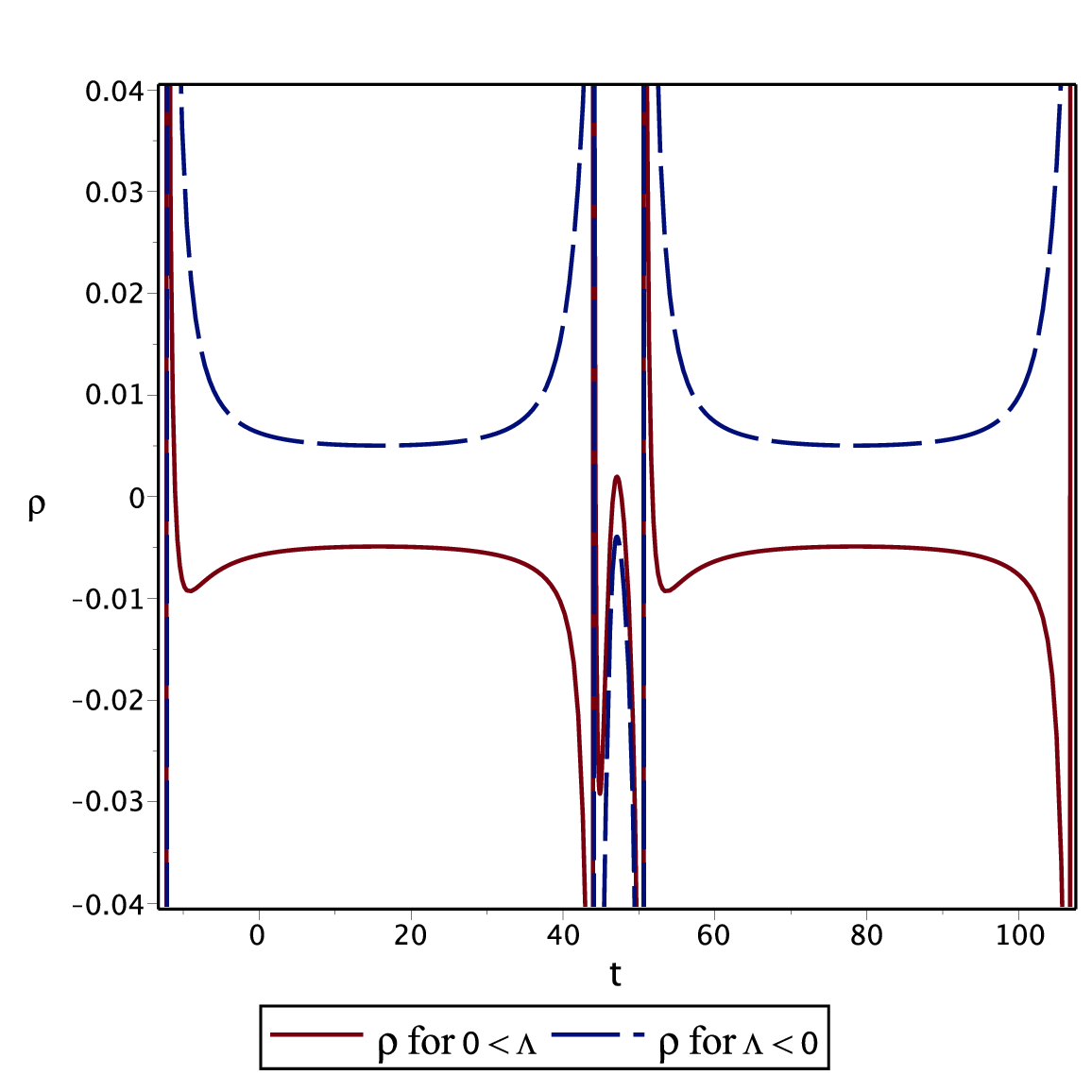}} 
		 \hspace{0.5cm}
		 \subfigure[$p$]{\label{F6}\includegraphics[width=0.3\textwidth]{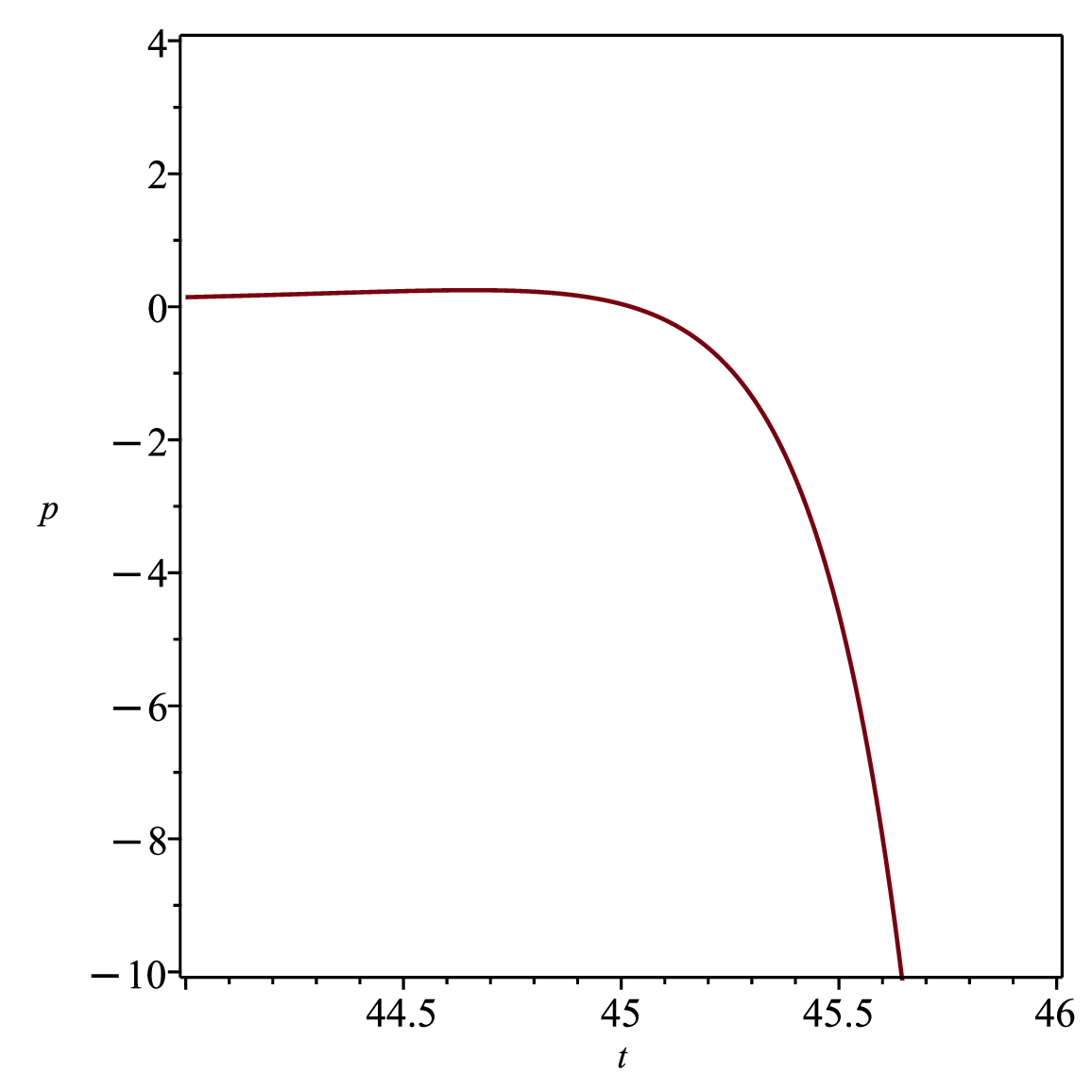}}\\
		\subfigure[$w$]{\label{F423}\includegraphics[width=0.3\textwidth]{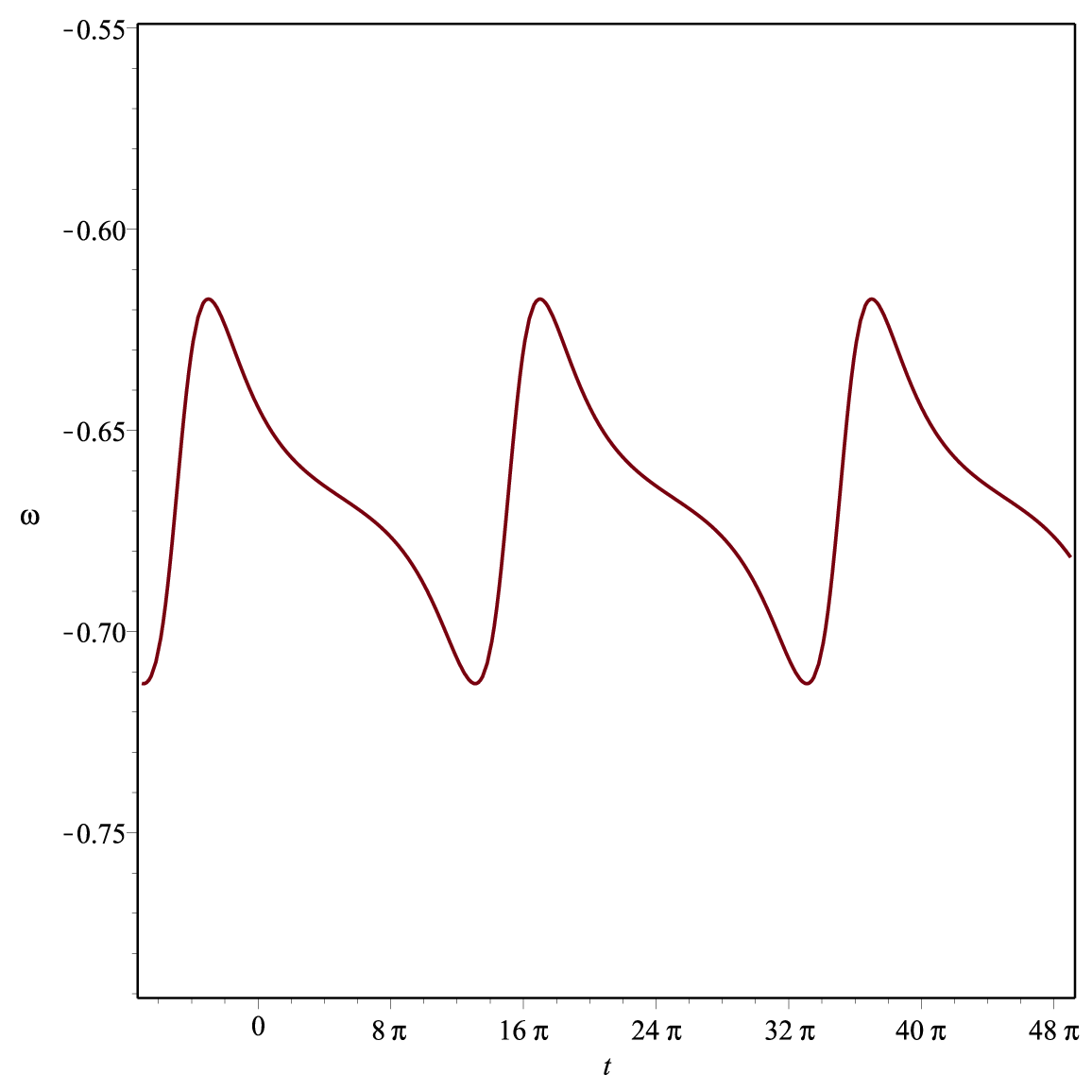}} 
		 \hspace{0.5cm}
			\subfigure[$q$]{\label{F3}\includegraphics[width=0.3\textwidth]{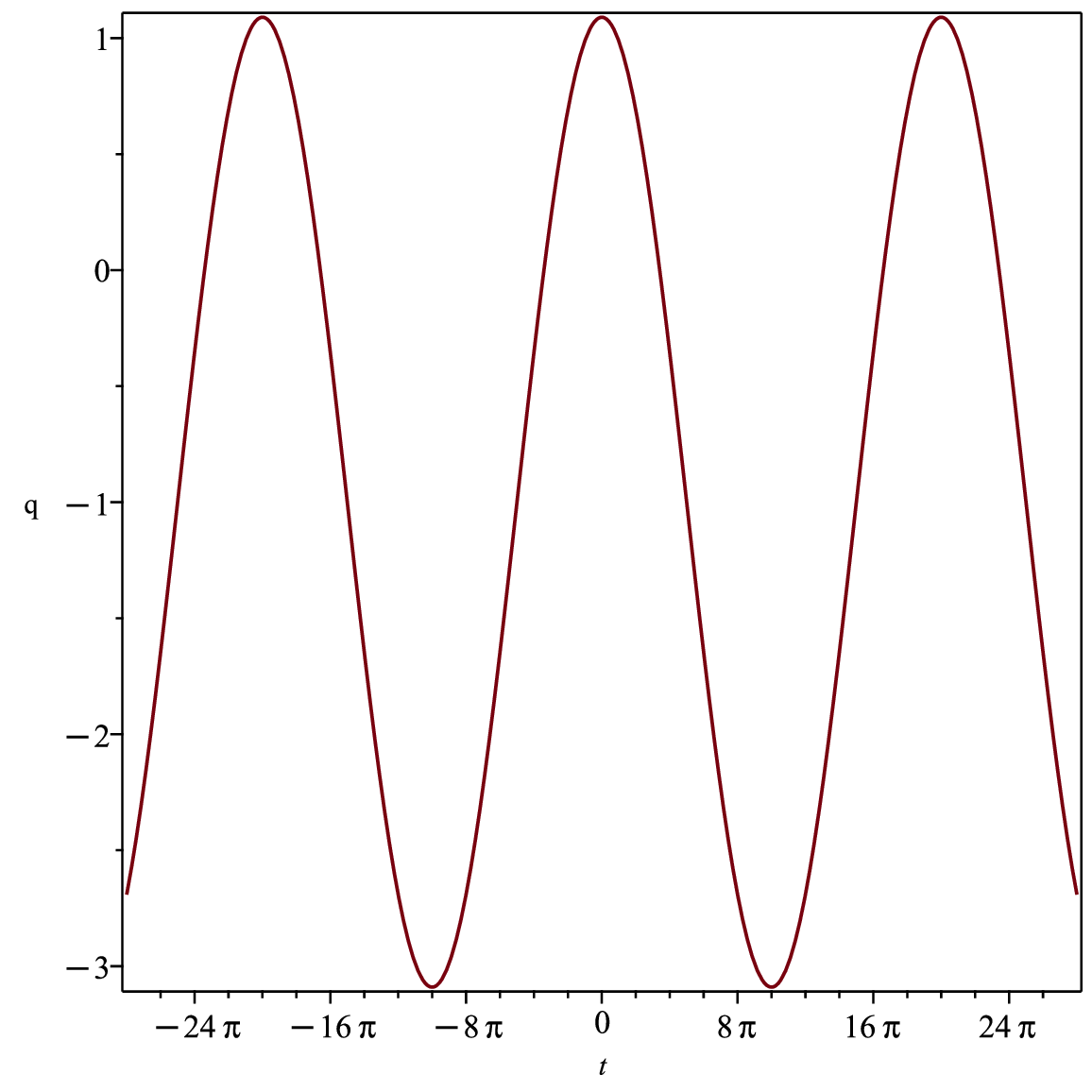}} 
  \caption{Evolution of $\Lambda$, $\rho$, $p$, $w$ and $q$ as a function of the cosmic time t: (a) The positive $\Lambda$ ansatz (formula (\ref{vary2})) with the new suggested one for negative $\Lambda$ (formula (\ref{vary3})). (b) $\rho$ for positive and negative $\Lambda$ cases. For $\Lambda < 0$, It remains positive for most of the evolution and gets negative within a small neighborhood of the bounce. Negative $\rho$ is a necessary component in most cyclic models that enables the transition from accelerated expansion to contraction and then a bounce. We find $\rho < 0$ for $\Lambda \geq 0$. (c) Sign flipping of pressure $p$. (d) The EoS parameter $w$ mainly lies in the quintessence region $ -0.75 < w< -0.6$ without crossing to the cosmological constant boundary $w=-1$. (e) Deceleration parameter $q$ lies in the range $-3 \leq q \lesssim 1$ . Cosmic transit is revealed by the sign flipping of $q$ and $p$. Here: $\alpha_0=1$, $\lambda_0=0.1$, $\beta_0=0.5$, $c=2.1$, $m=2.09$, $M=1$. }
  \label{fig:cassimir55}
\end{figure}

\quad Using the proposed ansatz (\ref{vary3}) we solve Eqs. (\ref{RW1}) and (\ref{RW2}) for the observationally supported flat case and get the following expressions for cosmic pressure, energy density and equation of state parameter
\begin{eqnarray}
p&=&-~\frac{ -2\lambda \Lambda a^2+3\lambda \dot{a}^2+9a \dot{a}\dot{\lambda}+3a\lambda \ddot{a}+3a^2\ddot{\lambda} }{24 \pi \lambda a^2},\\
\rho&=& \frac{ -\Lambda \lambda^2 a^2+3\lambda^2 \dot{a}^2+6\lambda a \dot{\lambda}\dot{a}+3a^2 \dot{\lambda}^2 }{8 \pi \lambda^2 a^2},\\
w&=&-\frac{\lambda}{3}~\frac{-2\lambda \Lambda a^2+3\lambda \dot{a}^2+9a \dot{a}\dot{\lambda}+3a\lambda \ddot{a}+3a^2\ddot{\lambda}}{-\Lambda \lambda^2 a^2+3\lambda^2 \dot{a}^2+6\lambda a \dot{\lambda}\dot{a}+3a^2 \dot{\lambda}^2 },
\end{eqnarray}
where the scale factor is given by the ansatz (\ref{a}), $\lambda^2=e^Q$ is the quantum potential $Q$ is given by Eq.~(\ref{conf}), and the dots denote derivatives with respect to time. The old positive and new negative $\Lambda$ ansätze have been plotted in Fig.~\ref{fig:cassimir55}(a). Figure~\ref{fig:cassimir55}(b) shows the energy density for both the positive $\Lambda$ solution using formula (\ref{vary2}), and the negative $\Lambda$ solution using the current suggested formula (\ref{vary3}). The figure suggests that the physical solution is the negative $\Lambda$ one, while positive $\Lambda$ leads to domination of negative energy density. The behavior of the physically acceptable $\rho$ also shows that, besides its positivity for most of the evolution, it can also reach negative values in a small neighborhood of the bounce. \par
\quad Negative energy density is proposed in some cyclic cosmological models as a necessary component for transition from accelerated expansion to contraction and then bounce \cite{negative1, negative2}. Potentially, it is driven by exotic components like negative phantom energy, negative cosmological constants, or other quantum effects. In Ref.~\cite{negative2}, a negative potential energy has been introduced rather than spatial curvature to cause the transition from expansion to contraction. They assumed a potential $V(\phi)$ such that it tends to zero rapidly as $\phi \rightarrow -\infty$, and it is negative for the intermediate $\phi$ before it rises to positive. Negative vacuum energy is then a different mechanism for getting an oscillating universe where it plays the role of a positive spatial curvature and leads to a collapse as the AdS behaviour is approached \cite{negative4}. A possible scenario has been considered in Ref.~\cite{negative3} as follows: as the kinetic energy density starts to exceed the decreasing positive potential energy density, the accelerated expansion era comes to an end and the decelerated expansion era starts. As the scalar field evolves down the potential, the potential energy density gets sufficiently negative so that the total energy density reaches zero.  Consequently, the expansion ($H>0$) stops and the contraction ($H<0$) starts. \par
\quad The bounce in cyclic models is often driven by a form of dark energy (DE) or other exotic matter with negative pressure, but the total energy density remains positive to allow for the rebound. Figure~\ref{fig:cassimir55}(c) shows a sign change of pressure in each cycle. In cosmic history, cosmic pressure is supposed to be positive in the early deceleration time and negative during the late-time acceleration. According to the standard cosmological model, the early universe (z$\rightarrow \infty$) is assumed to be dominated by positive pressure while negative pressure dominates in the far accelerating epoch. Hence, such evolution of pressure from positive to negative can play a key role in explaining the deceleration-acceleration cosmic transit.\par

\quad The deceleration parameter $q=\frac{1}{2}(1+3~w)$ (Fig.~\ref{fig:cassimir55}(e)) lies in the range $-3 \leq q \lesssim 1$ which includes radiation-dominated era ($q=1$), matter-dominated era ($q=\frac{1}{2}$), cosmic transit ($q=0$), dark energy ($q=-1$), quintessence ($-1<q<0$) and phantom energy ($q<-1$). For FLRW metric, the redshift is related to the scale factor by $a=\frac{1}{1+z}$. Using this relation with Eq.~(\ref{a}), cosmic time can be expressed as a function of $z$ as  
\begin{equation} \label{t}
t=\frac{2}{k} \arctan \left\{\frac{1}{c}\left(\sqrt{c^2-m^2}-m\right)\tan \left[\frac{1}{2}(\sqrt{c^2-m^2}) \ln\left(\frac{1}{A(1+z)}\right)\right]\right\}.
\end{equation}
\quad Cosmic transit happens when the deceleration parameter vanishes $q=0$ (i.e., $\ddot a =0$). If the sign flipping of $q$ occurs at $z=0.064$ for $m=1.55$ as some observations suggest \cite{4,5}, then the above equation for cosmic time gives $t_{q=0}\approx 6-9$ Gyr for $A \approx 0.001$. \par
\quad The periodic varying $w$ shows a quintessence-dominated universe associated with no quintom behavior. In general, the equation of state parameter for the quintessence field lies in the range $ w \in [-1,1]$. When $ w \approx -1$, the quintessence field acts like a cosmological constant. Quintom models have got the advantage of being able to describe the crossing of the phantom divide. In quintessence and phantom models, the varying $w$ always remains on the same side of the phantom line $ w > -1$ for quintessence and $ w < -1$ for phantom. A bouncing universe dominated by quintom matter has been investigated in Ref.~\cite{dd2}. 
\subsection{Positive and vanishing $\Lambda$}
\quad Using the ansatz for the negative time-varying $\Lambda$ in Eq.~(\ref{vary3}), we have obtained the following expression for energy density
\begin{equation} \label{rhoo}
\rho=\frac{ -\Lambda \lambda^2 a^2+3\lambda^2 \dot{a}^2+6\lambda a \dot{\lambda}\dot{a}+3a^2 \dot{\lambda}^2 }{8 \pi \lambda^2 a^2} >0, ~~~ \forall ~ t \notin [t_b-\Delta t , t_b+\Delta t] .
\end{equation}
\quad As we have indicated in Sec. \ref{secprho}, this energy density is positive for most of the evolution except in a small neighborhood of the bounce $t_b$. On the contrary, replacing $-\Lambda \rightarrow \Lambda$ in Eq.~(\ref{rhoo}) leads to domination of negative energy density through the cycle. Same undesired result for replacing $-\Lambda \rightarrow \Lambda$ is also obtained by setting $\Lambda$ to zero in the above expression. Based on the sign of $\rho$, we consider the physically acceptable solution is the one with $\Lambda < 0$ and $\rho > 0$. Setting $\rho > 0$ in Eq.~(\ref{rhoo}) directly implies the following inequality for the cosmological constant satisfied for all $t$ except within a finite interval containing the bounce
\begin{equation} 
\Lambda < 3 \left(\frac{\dot{\lambda}}{\lambda}\right)^2 + 3 \left(\frac{\dot{a}}{a}\right)^2 + 6 \left(\frac{\dot{\lambda}}{\lambda}\right)\left(\frac{\dot{a}}{a}\right) . 
\end{equation} 
\quad Negative energy density or negative pressure is specifically required in cyclic models to trigger transition from contraction to expansion (the bounce) \cite{negative1}, often in the form of a negative potential, phantom energy, or curvature. Such negative energy component does not necessarily dominate the entire cycle. While dark energy (negative pressure component) accelerates expansion during a part of a cycle, the negative energy component is responsible for the bounce. The null energy condition is temporarily violated near the bounce. While many proposals for the bounce have been suggested in different contexts, the exact nature of the bounce remains unclear. 

\quad An earlier study has been done in Ref.~\cite{kamel} in the context of conformal Bohm-de Broglie gravity (same modified gravity theory of the present study) but for $\Lambda > 0$. Although both studies predict a future quintessence-dominated universe, the current one for negative $\Lambda$ seems to provide a more complete description for cosmic evolution and dark energy. The three positive-$\Lambda$ solutions in Ref.~ \cite{kamel} respectively are the hyperbolic $a(t)= A\sinh^n(b t)$ for $0 < n < 1$, the logamediate Inflation $a(t)=e^{B\ln(t)^{\alpha}}$, and the hybrid $a(t)=a_1 t^{\alpha_1} e^{\beta t}$ where $a_1>0$, $\alpha_1 \geq 0$ and $\beta \geq 0$ are constants. Despite the presence of cosmic transit in the three models, the pressure consistently remains negative throughout cosmic evolution. Furthermore, the equation of state parameter's evolution invariably contains a singularity, making it not suitable for a complete description of dark energy evolution. Such undesired cosmic features in pressure $p$ and equation of state parameter $w$ have been avoided in the present negative-$\Lambda$ model where $p$ reveals a positive-negative sign change in corresponding to the positive-negative sign change of the deceleration parameter $q$, and $w$ reveals no improper behavior or singularities. The three empirical forms of the scale factor used in Ref.~\cite{kamel} are shown to be observationally consistent and led to excellent cosmic behavior for all parameters in other modified gravity theories. 

\section{Kinetic and potential terms} \label{seckv}
\quad The energy density and pressure for a universe filled with a scalar field $\phi$ are related by $\rho_{\phi}=K+V(\phi)$ and $p_{\phi}=K-V(\phi)$, where, $K=\frac{1}{2}\dot{\phi}^2$ is the kinetic term and $V=V(\phi)$ is the potential. We then find 
\begin{eqnarray}
K&=&\frac{1}{2}(\rho_{\phi}+p_{\phi})\\   \nonumber
&=& -\frac{1}{48 a^2\lambda^2 \pi} \left( \Lambda a^2\lambda^2 + 3a^2\lambda \ddot{\lambda}-9a^2 \dot{\lambda}^2+3a\lambda^2\ddot{a}-9a \dot{a} \lambda \dot{\lambda}-6\lambda^2 \dot{a}^2 \right),
\end{eqnarray}
\begin{eqnarray}
V&=&\frac{1}{2}(\rho_{\phi}-p_{\phi}) \\   \nonumber
&=& -\frac{1}{48 a^2\lambda^2 \pi}    \left( 5\Lambda a^2\lambda^2 - 3a^2\lambda \ddot{\lambda}-9a^2 \dot{\lambda}^2-3a\lambda^2\ddot{a}-27a \dot{a} \lambda \dot{\lambda}-12\lambda^2 \dot{a}^2 \right).
\end{eqnarray} 
In terms of the equation of state parameter  $w_{\phi}$ we have $\dot{\phi}^2=(1+ w_{\phi}) \rho_{\phi}$ with $p_{\phi}=w_{\phi} \rho_{\phi}$. $w_{\phi}$ can be defined as
\begin{equation} \label{scc}
w_{\phi}= \frac{\frac{1}{2}\dot{\phi}^2-V}{\frac{1}{2}\dot{\phi}^2+V} \simeq -1 ~~~ \text{if} ~~~\frac{1}{2}\dot{\phi}^2 \ll V.
\end{equation}
Both $\rho_{\phi}$ and $p_{\phi}$ satisfy the conservation equation
\begin{equation} \label{cons}
\dot{\rho_{\phi}}+3H(\rho_{\phi}+p_{\phi})=0.
\end{equation}
The evolution equation for the scalar field is
\begin{equation} \label{evol}
\ddot{\phi}+3H\dot{\phi}+V_{,\phi}=0, 
\end{equation}
where $V_{,\phi}=dV/d \phi$. The scalar field $\phi(t)$ is given by
\begin{eqnarray} \label{scalarf}
\phi(t)&=& \int_0^t \pm ~\sqrt{\rho_{\phi}+p_{\phi} } ~dt +\phi_0, \\  \nonumber
&=& \int_0^t \pm ~\frac{1}{2a\lambda}\sqrt{ \frac{1}{6 \pi} \left( \Lambda a^2\lambda^2 + 3a^2\lambda \ddot{\lambda}-9a^2 \dot{\lambda}^2+3a\lambda^2\ddot{a}-9a \dot{a} \lambda \dot{\lambda}-6\lambda^2 \dot{a}^2 \right)} ~dt +\phi_0,
\end{eqnarray}
where $\phi_0=\phi(t_0)=0$ can be set to zero without loss of generality. This integration can be numerically evaluated and plotted taking into account the domain of the function $\sqrt{\rho +p}$. To plot $V(\phi)$ and see the AdS minima directly, we need to get $t(\phi)$ and substitute in the expression for $V(t)$ which is too complicated for the present model. Since it’s hard to invert $\phi(t)$ to get an analytic form of $V(\phi)$, we plot $V(t)$ which shows how the potential energy evolves with cosmic time (the value of the potential energy of the scalar field at a given moment in cosmic history). Plotting $V(\phi)$ shows the shape and the location ($\phi$-value) of the minimum. However, if $V(t)$ just becomes negative during evolution it is not enough to conclude AdS. When a scalar field with a negative potential or a negative cosmological constant is introduced into a non-singular cosmological model, it is widely believed that a cyclic evolution will be easily obtained, leading to a recollapse at some point in the evolution \cite{loop}. Figure~(\ref{fig:camir55}) shows that the kinetic term and scalar potential are always positive, The sum $V+Q$ is negative. The positive kinetic term indicates the absence of ghost-like behavior.\par
\quad Because the potential in the present work is not reconstructed explicitly as $V(\phi)$ but inferred indirectly from the background evolution, statements about its form and physical properties remain indicative rather than definitive. A complete and unambiguous physical interpretation would require an explicit reconstruction of the scalar field dynamics, including the functional form of $V(\phi)$, as well as an analysis of stability and perturbations. Such an extension would be a natural and important direction for future work. \par

\quad If the scalar field inside a bubble universe lands at the AdS minimum of its effective potential $V(\phi)$, this universe will collapse and bounce again \cite{etern}. Hence, the universe collapses as the scalar field is trapped in AdS potential well, while it expands for $V(\phi)>0$ (accelerated expansion). The universe will reach a dS state When the scalar field lands at the dS minimum. 
Negative potentials have attracted attention in particle physics and cosmology since the string theory prediction of AdS spaces. They appear in super-gravity as well as in cyclic and ekpyrotic cosmological models \cite{cyc,ekp}. Explaining the cosmic scale using a high energy scale, like the super-symmetry breaking scale or the electroweak scale, is also a result of negative potentials \cite{scale}. A discussion of scalar field cosmology with negative potentials has been presented in Ref.~\cite{linde}. Negative energy densities also play a role in FLRW cosmological models which has been investigated in Ref.~\cite{-ve}. In terms of the scale factor, the total energy density can be expressed as the sum of two power series
\begin{equation}
\rho=\sum_{n=-\infty}^{\infty}\rho_n^+a^{-n}+\sum_{m=-\infty}^{\infty}\rho_m^-a^{-m},
\end{equation}
where $\rho_n^+$ is the normal positive $\rho$, and $\rho_m^-$ is the negative cosmological energy density. 
\section{Energy conditions} \label{conditions}
\quad Violations of energy conditions (especially Null and Strong conditions) are common near bounces in cyclic models in both GR and modified gravity frameworks, they are often essential to enable the bounce itself. To go from contraction to expansion through a non-singular bounce in traditional bouncing cosmologies, the standard GR Friedmann equation demands that the Hubble parameter passes through zero. That typically requires temporarily violating the Null Energy Condition (NEC) $\rho + p\geq 0$ near the bounce (e.g., with ghost condensates or phantom matter). This NEC violation usually implies violations of other point-wise energy conditions, like the Strong Energy Condition (SEC), in the vicinity of the bounce \cite{Bo1,Bo2}. While most non-singular cosmological models require the NEC violation, it's highly preferable to avoid such violation if possible. \par
\quad Among all Energy Conditions (ECs), The NEC is considered the most fundamental and its violation automatically leads to the violation of all other point-wise ECs. A classical non-singular model where the NEC is not violated has been introduced in Ref.~\cite{necv}. A non-singular bouncing  model with flat scalar potential and positive curvature has been presented in Ref.~\cite{necv3}.
\begin{figure}[t]
  \centering         
				\subfigure[$K$]{\label{020}\includegraphics[width=0.3\textwidth]{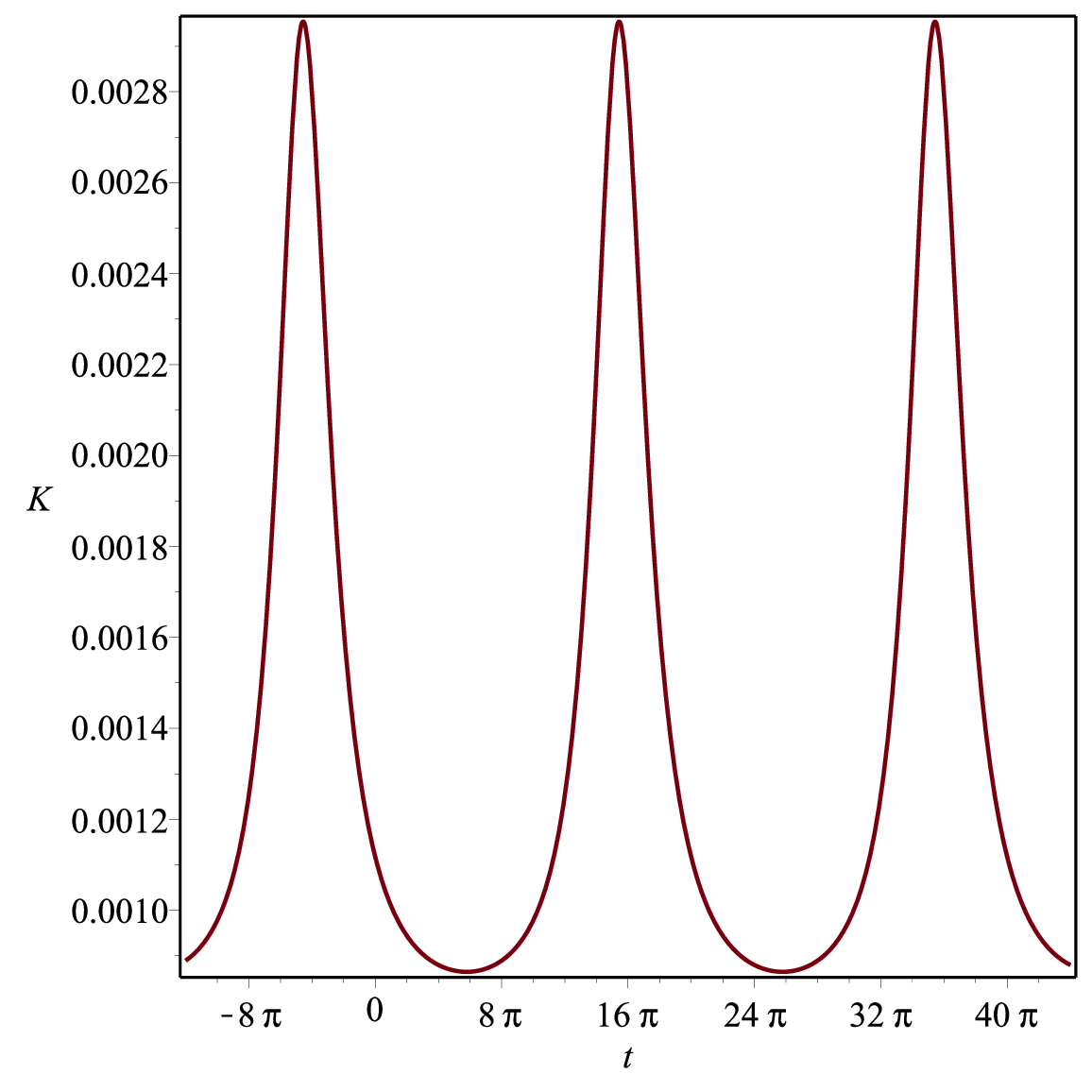}}
				     \hspace{0.5cm}
			\subfigure[$V$]{\label{021}\includegraphics[width=0.3\textwidth]{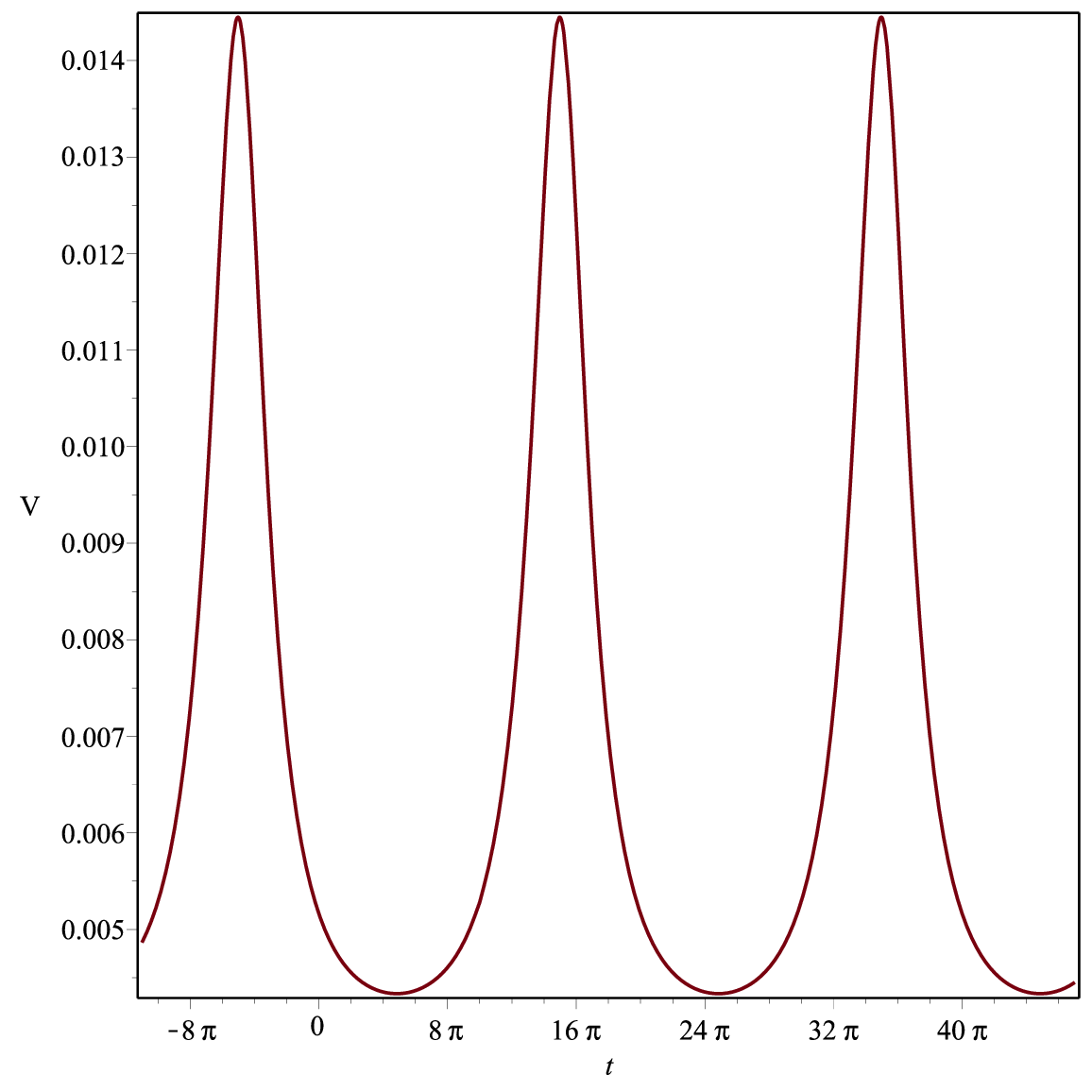}}
			     \hspace{0.5cm}
	\subfigure[$Q$]{\label{F42733}\includegraphics[width=0.3\textwidth]{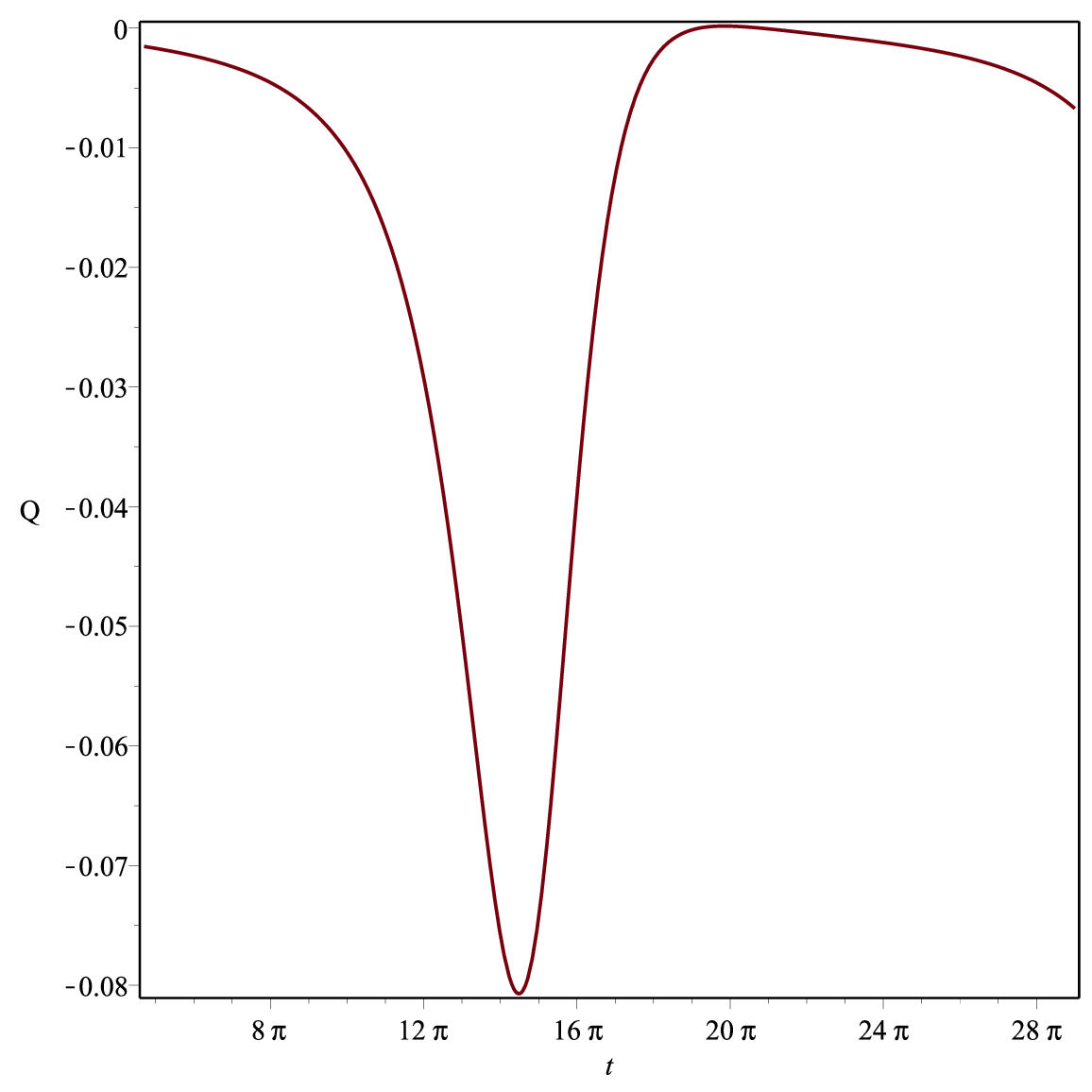}}\\
	\subfigure[$V+Q$]{\label{F4273}\includegraphics[width=0.3\textwidth]{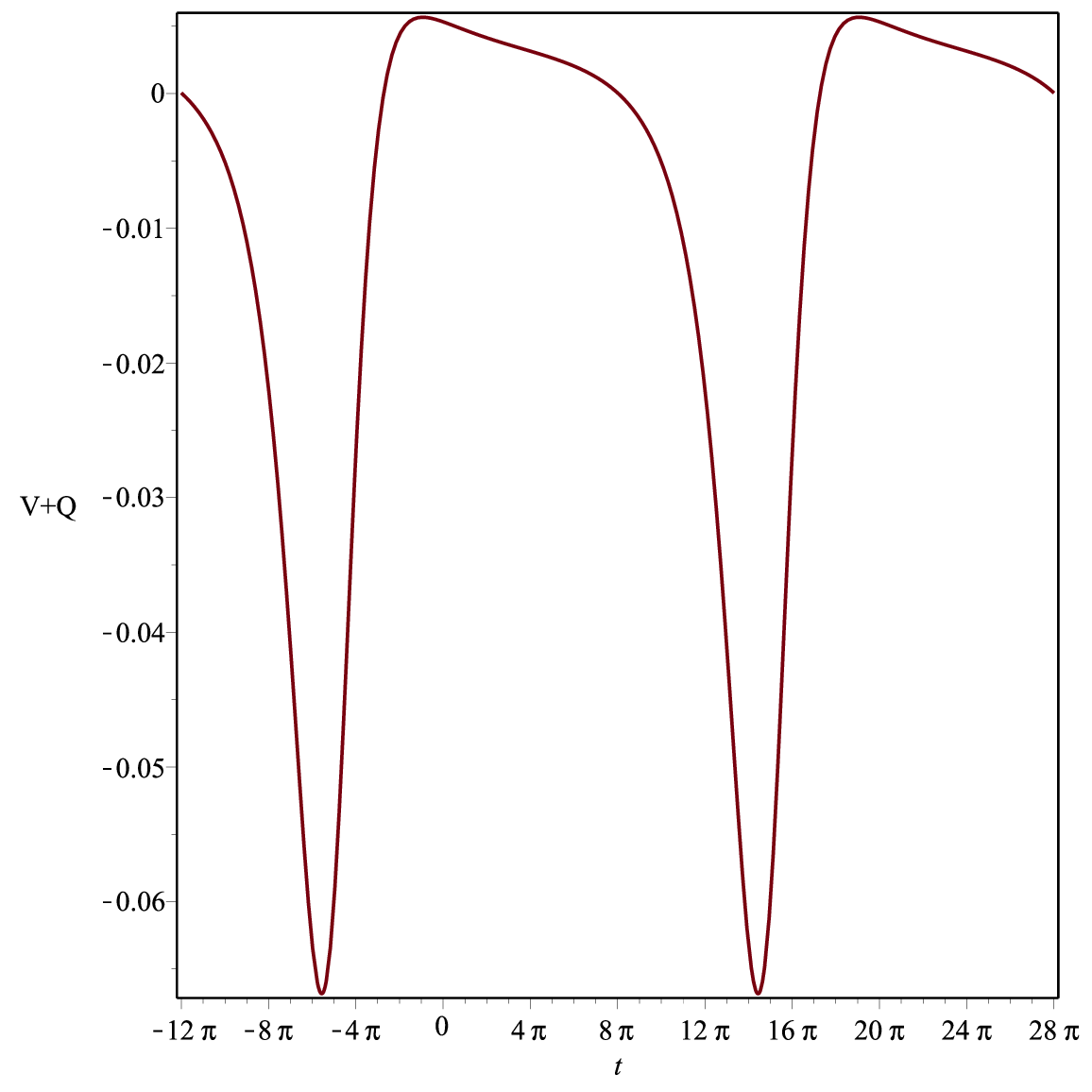}}
	 \hspace{0.5cm}
		\subfigure[Linear ECs]{\label{r0}\includegraphics[width=0.3\textwidth]{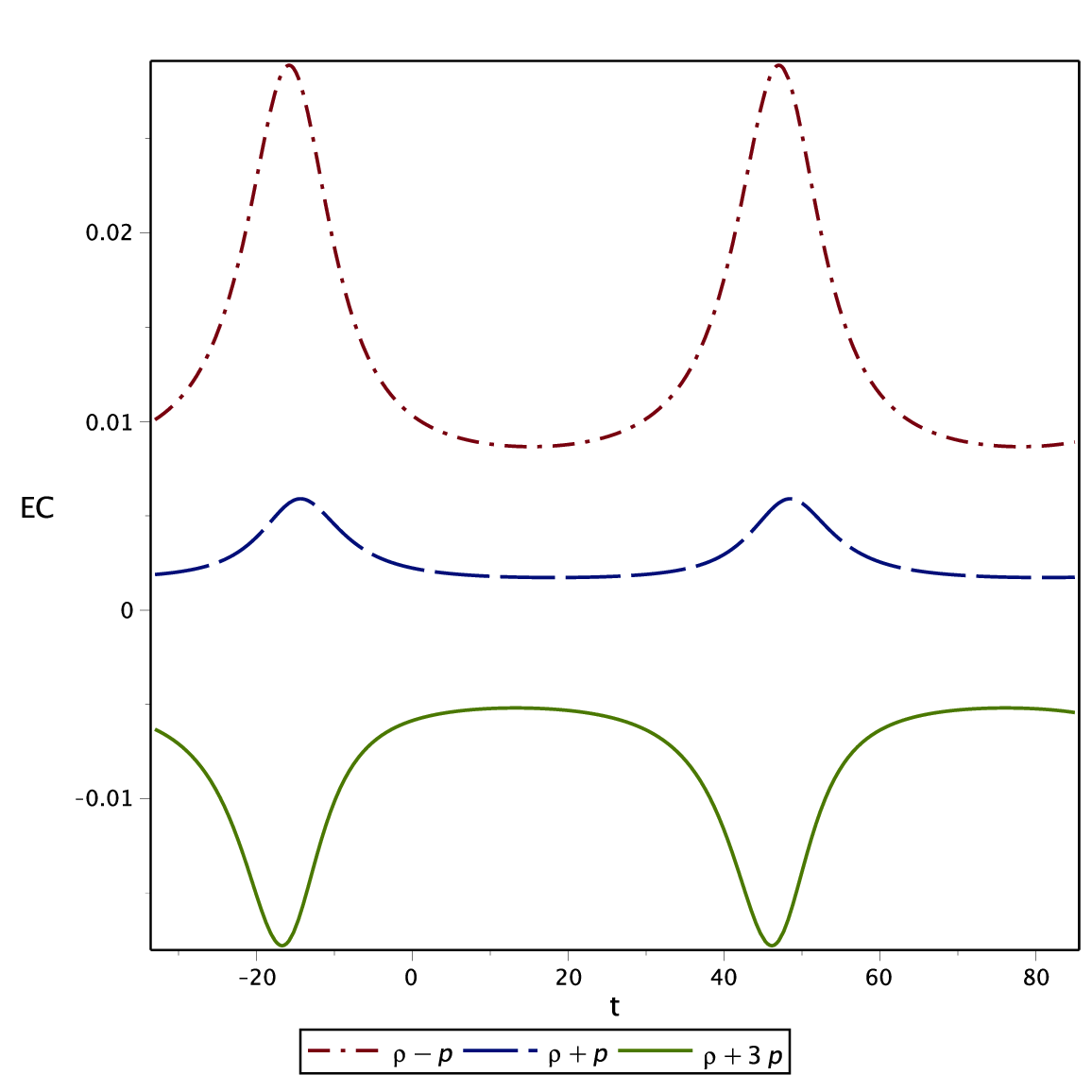}}
		 \hspace{0.5cm}
			\subfigure[Nonlinear ECs]{\label{t21}\includegraphics[width=0.3\textwidth]{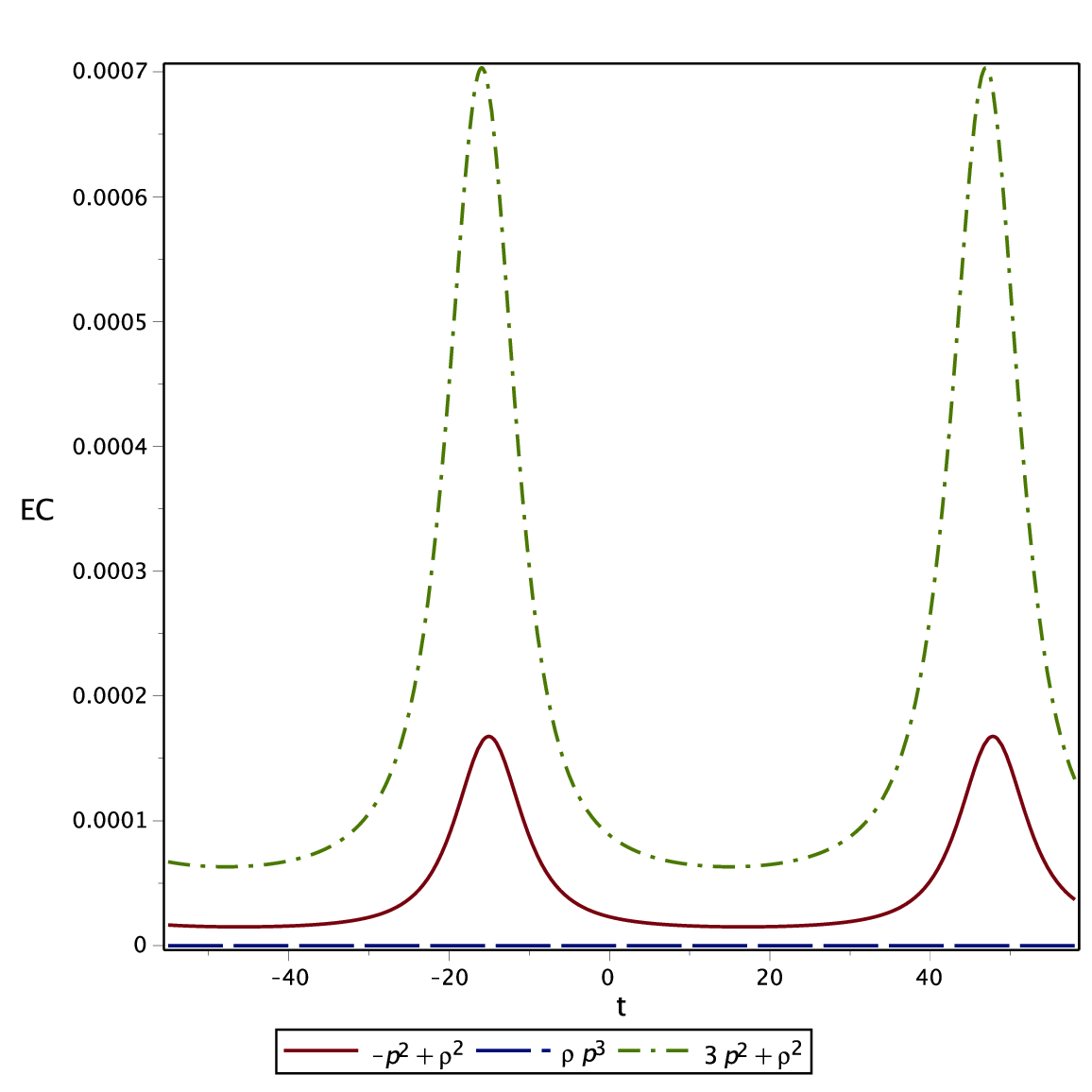}}
  \caption{$K$, $V$, $Q$, and the sum $V+Q$ as functions of the cosmic time $t$: $V > 0$, $K > 0$ and $V+Q < 0$. NEC and DEC are always satisfied, only SEC is violated as expected. Nonlinear conditions FEC, DETEC and TOSEC are satisfied all the time. Avoiding NEC violation is highly preferable although it is required in most non-singular cosmological models.}
  \label{fig:camir55}
\end{figure}
The classical linear energy conditions \cite{ec11,ec12}, and the new nonlinear energy conditions (if semi-classical quantum effects exists) \cite{ec,FEC1, FEC2} can be used to test the physical acceptability of the current model. The linear conditions are the null $\rho + p\geq 0$; weak $\rho \geq 0$, $\rho + p\geq 0$; strong $\rho + 3p\geq 0$ and dominant $\rho \geq \left|p\right|$. Although they work extremely well in classical gravity, they get violated by semi-classical quantum effects and can not be generally valid \cite{ec2}. The highly restrictive SEC states that gravity should always be attractive which does not apply when describing the present cosmic acceleration phase and throughout inflation. The NEC is not violated for the present model. Due to the existence of quantum effects represented in the quantum potential, we consider the validity of the following nonlinear conditions (i) The flux energy condition (FEC): $\rho^2 \geq p_i^2$ \cite{FEC1, FEC2}, first presented in Ref.~\cite{FEC1}. (ii) The determinant energy condition (DETEC): $ \rho . \Pi p_i \geq 0$ \cite{ec}. (ii) The trace-of-square energy condition (TOSEC): $\rho^2 + \sum p_i^2 \geq 0$ \cite{ec}. Figure~\ref{fig:camir55} shows that they are all satisfied through each cycle.
\begin{table}
\caption{Comparison between the existence of negative $\Lambda$ in some modified gravity frameworks and the current model.}
\begin{ruledtabular}
\begin{tabular}{p{5cm}@{\hspace{9pt}}p{5cm}@{\hspace{9pt}}p{5cm}@{\hspace{9pt}}}
   Modified gravity framework & Where $\Lambda <0$ appears & Main role \\
\hline
RS model (Brane-world) \cite{RS,nas333}& Fundamental bulk cosmological constant $(\Lambda_5 < 0)$ & Produces AdS bulk geometry, localizes gravity on the brane. \\
Scalar–tensor gravity (ekpyrotic/cyclic models) \cite{cyc,ekp} & Scalar potential $V(\phi) < 0$ (effective $\Lambda$) & enables cyclic or bouncing cosmologies.  \\
	AdS/CFT-motivated cosmology \cite{ah} & Fundamental $\Lambda < 0$ & Enables holographic dual descriptions of cosmology and quantum gravity effects.  \\
Hořava–Lifshitz gravity \cite{hl}& Bare $\Lambda$ & Improves UV consistency and often required for detailed balance.  \\
Gauss–Bonnet/Lovelock gravity \cite{ma,ma2,ma3} & Bare or effective $\Lambda < 0$ & leads to asymptotically AdS 5D black holes, which can have nontrivial topologies, stabilizes dS solutions in higher-dimensions.  \\
Graduated Dark Energy \cite{GDE} & Bare $\Lambda < 0$ & provides the true vacuum structure, produces late-time acceleration without eternal dS expansion, provides a better fit for observational data compared to \(\Lambda \)CDM.  \\
Conformal Bohm-de Broglie gravity (present work)& time-varying $\Lambda(H) < 0$  & leads to cyclic evolution with no violation of NEC, cosmic transit, quintessence-dominated universe. Negative $\rho$ dominates for $\Lambda \geq 0$.
\end{tabular}
\end{ruledtabular}
\end{table}

\section{Matter Bounce Scenario } \label{bounce1}
\quad Among numerous bouncing models that have been developed, the Matter Bounce Scenario (MBC) has attracted a special attentioan where it can generate a power spectrum of primordial curvature perturbations that is almost scale-invariant  \cite{bounc11,bounc12}. The universe in MBS is nearly matter-dominated during the initial stages of the contracting phase and slowly transitions towards a bounce. Every region of the universe is assumed to be in causal contact at the bounce implying that the horizon problem does not arise. Subsequently, a standard phase of expansion begins in agreement with the evolution predicted by the standard cosmological model. Several conceptual aspects of this scenario have been discussed in Ref.~\cite{bounc5}. The scale factor for a variant non-singular bounce is given by \cite{bounc5} 
\begin{equation} \label{scalef}
a(t)=\left(A t^2+1\right)^n
\end{equation}
The MBC can be explored for $n=\frac{1}{3}$. The deceleration and Hubble parameters are written as 
\begin{equation} \label{q15}
q(t)=-\frac{\ddot{a}a}{\dot{a}^2}=-\frac{(2n-1)A t^2+1}{2n A t^2}  \;\;\;\;\;\;\;,  \;\;\;\;\;\;\; H(t)= \frac{2n A t}{ A t^2+1}
\end{equation}
Utilizing formula (\ref{scalef}) for $n=\frac{1}{3}$ (MBS) in the suggested expression for negative $\Lambda$ (formula \ref{vary3}), and in the previously obtained solution for $\rho$, $p$ and $w$
\begin{eqnarray}
p&=&-~\frac{ -2\lambda \Lambda a^2+3\lambda \dot{a}^2+9a \dot{a}\dot{\lambda}+3a\lambda \ddot{a}+3a^2\ddot{\lambda} }{24 \pi \lambda a^2},\\
\rho&=& \frac{ -\Lambda \lambda^2 a^2+3\lambda^2 \dot{a}^2+6\lambda a \dot{\lambda}\dot{a}+3a^2 \dot{\lambda}^2 }{8 \pi \lambda^2 a^2},\\
w&=&-\frac{\lambda}{3}~\frac{-2\lambda \Lambda a^2+3\lambda \dot{a}^2+9a \dot{a}\dot{\lambda}+3a\lambda \ddot{a}+3a^2\ddot{\lambda}}{-\Lambda \lambda^2 a^2+3\lambda^2 \dot{a}^2+6\lambda a \dot{\lambda}\dot{a}+3a^2 \dot{\lambda}^2 },
\end{eqnarray}
\begin{figure}[t]
  \centering         
	\subfigure[$\Lambda$]{\label{pooi}\includegraphics[width=0.3\textwidth]{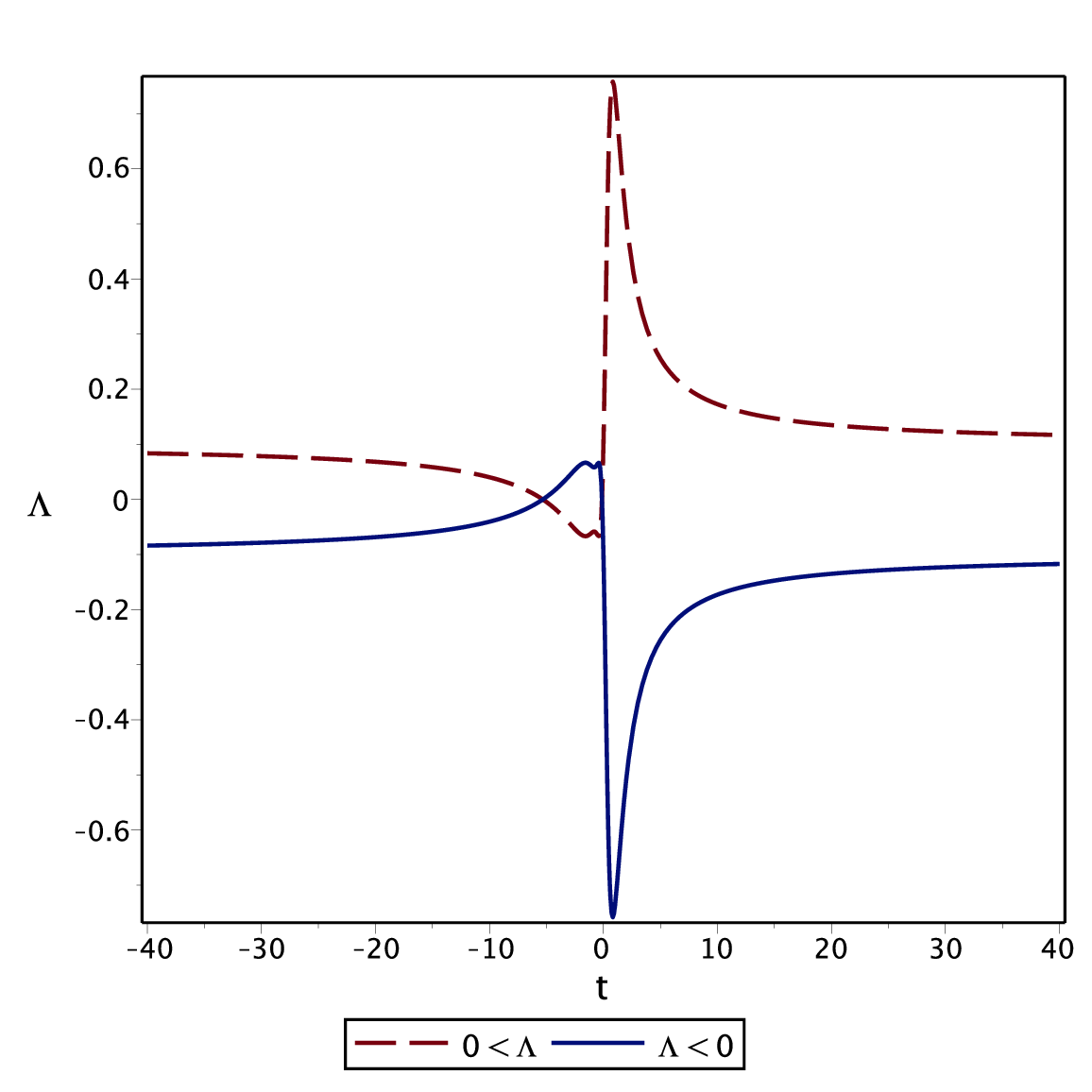}}
	 \hspace{0.5cm}
	  \subfigure[$\rho >0$ for $\Lambda < 0$]{\label{F6pp3}\includegraphics[width=0.3\textwidth]{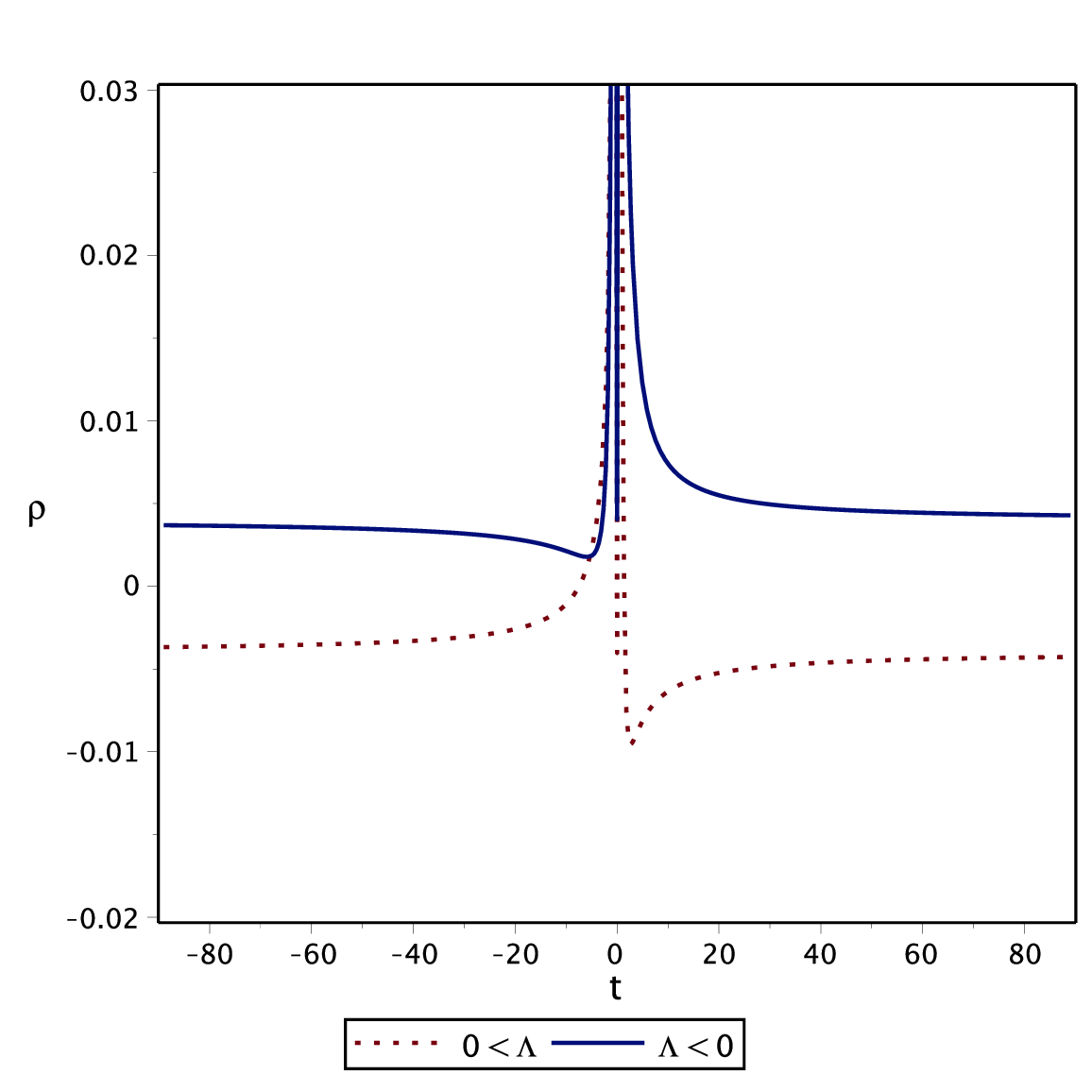}} 
		 \hspace{0.5cm}
		\subfigure[$w$]{\label{F4p23}\includegraphics[width=0.3\textwidth]{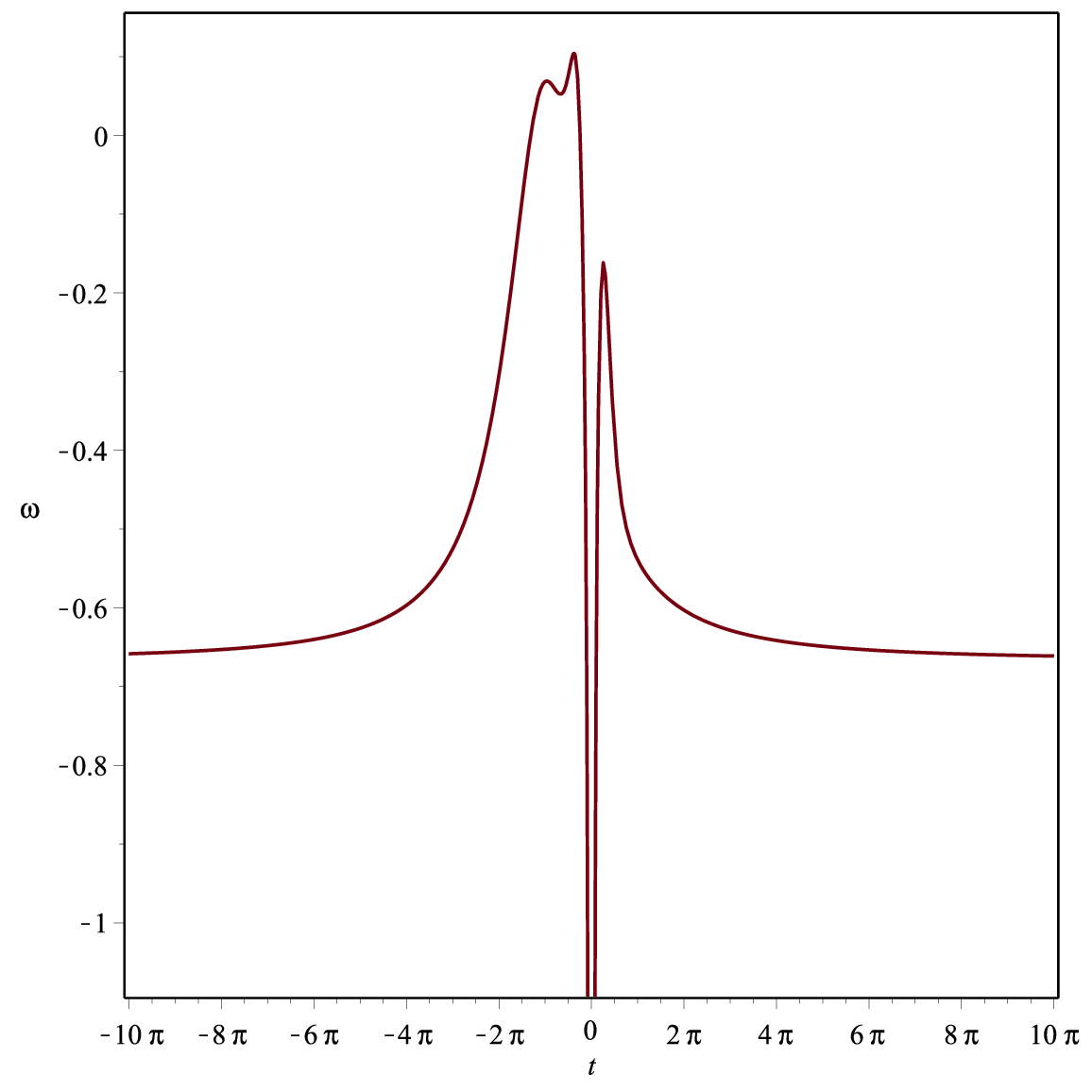}} \\
			\subfigure[$q$]{\label{Fp3}\includegraphics[width=0.3\textwidth]{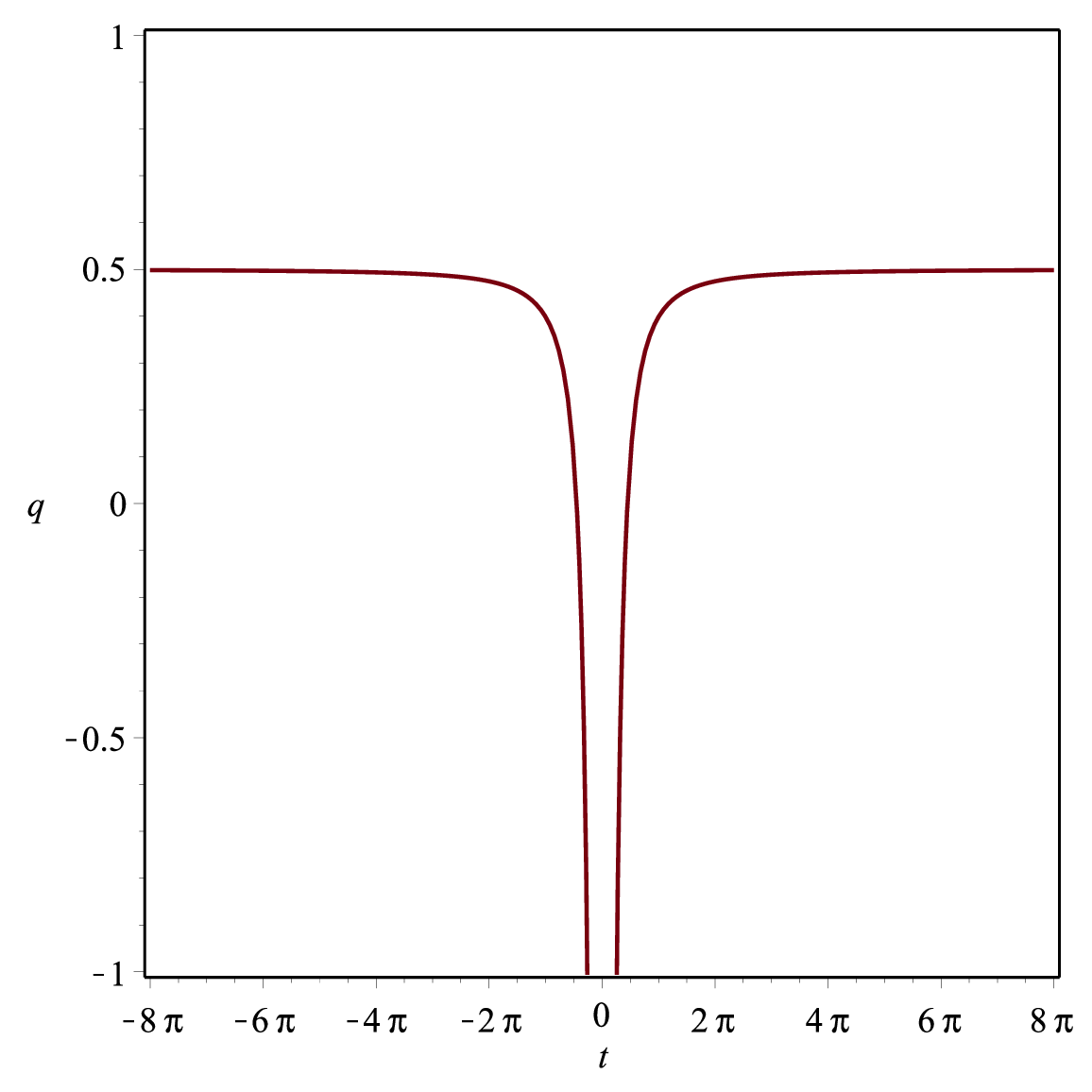}} 
			 \hspace{0.5cm}
			\subfigure[]{\label{po2oi}\includegraphics[width=0.3\textwidth]{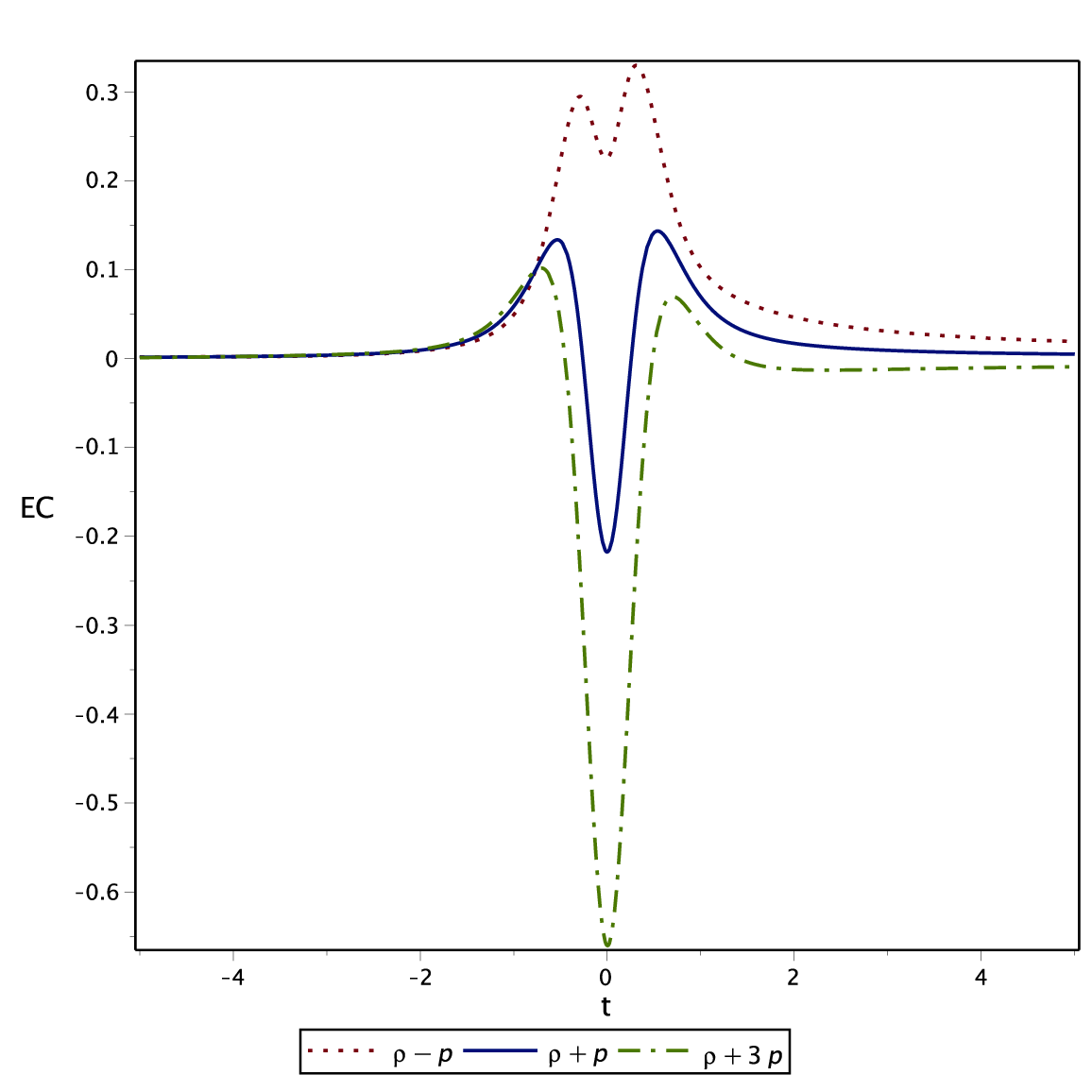}}
 \hspace{0.5cm}
				\subfigure[]{\label{p0oi}\includegraphics[width=0.3\textwidth]{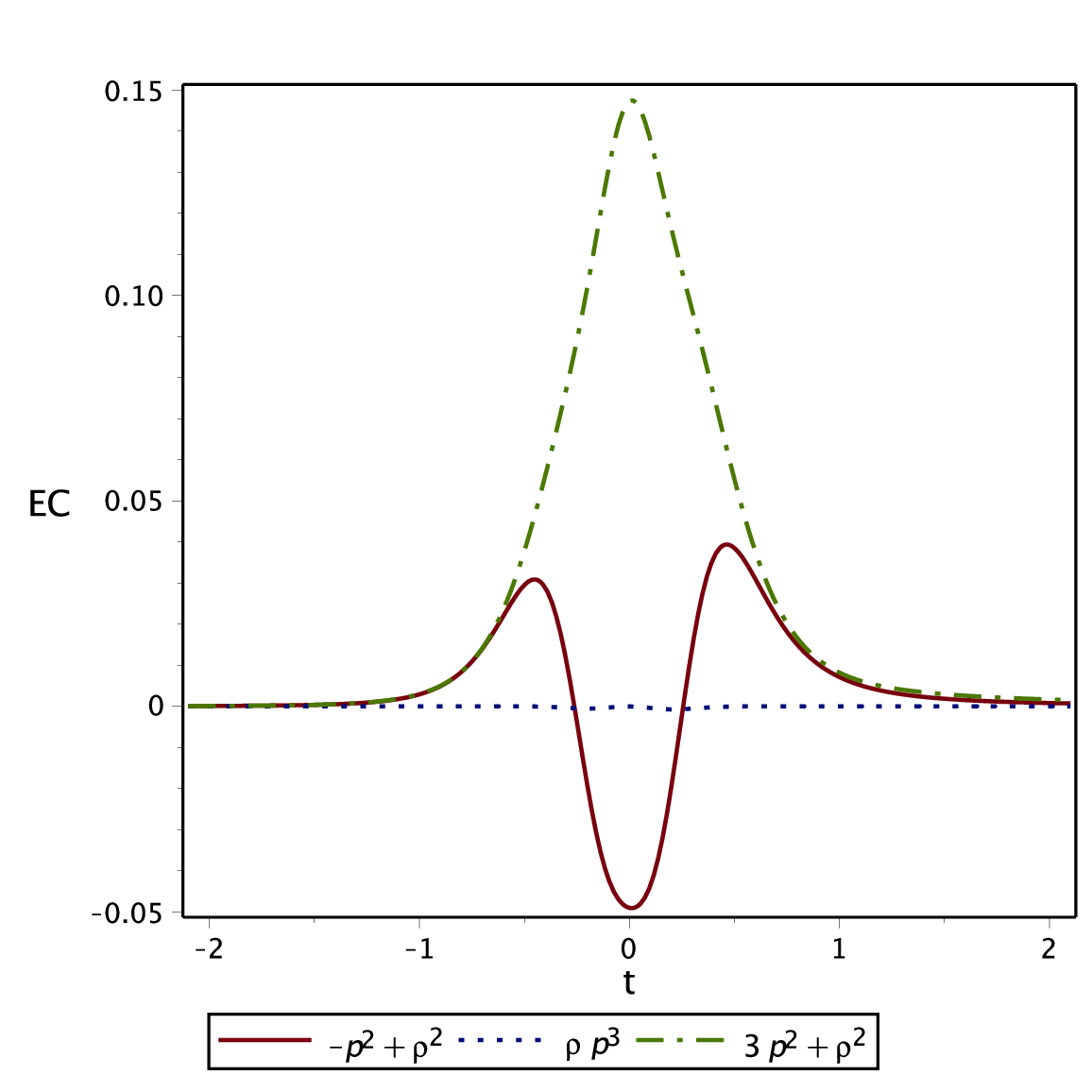}}
  \caption{Evolution of $\Lambda$, $\rho$, $p$, $w$ and $q$ as a function of the cosmic time t for MBS: (a) The positive $\Lambda$ ansatz (formula (\ref{vary2})) with the new proposed one for negative $\Lambda$ (formula (\ref{vary3})). (b) $\rho > 0$ only for the negative $\Lambda$ ansatz. (c) The EoS parameter $w$ mainly lies in the region $ -0.75 < w< 0.2$ with with a crossing to the cosmological constant boundary $w=-1$ (quintom behavior) only near the bounce. (d) Cosmic transit is revealed by the sign flipping of $q$ and $p$. (e) Linear ECs, The NEC gets violated only at the bounce. (f) Non-linear ECs . Here: $\alpha_0=1$, $\lambda_0=0.1$, $\beta_0=0.5$, $A=1.5$.}
  \label{cair55}
\end{figure}
\quad we can compare the behavior of energy density in the two cases of $0 < \Lambda$ (formula (\ref{vary2})) and $0 > \Lambda$. In Fig.~\ref{cair55}(a), the old positive and new negative $\Lambda$ ansätze have been plotted together for the MBS scale factor. Figure~\ref{cair55}(b) shows the energy density evolution in the two cases of the positive $\Lambda$ solution using formula (\ref{vary2}), and the negative $\Lambda$ solution using the new suggested formula (\ref{vary3}). Again, the physically acceptable solution is the negative $\Lambda$ one, while positive $\Lambda$ leads to domination of negative energy density. The solution reveals cosmic transit through the sign flipping in the evolution of pressure $p$ and deceleration parameter $q$. The time-varying $w$ shows a quintessence-dominated universe associated with quintom behavior only near the bounce where the crossing of the phantom divide line occurs. The NEC is satisfied but gets violated at the bounce.

\section{Conclusions} \label{conclusion}

\quad The laws of physics are generally not scale-invariant. As a consequence, the implications of the scale-invariance hypothesis to nature requires a careful examination. Accordingly, the cosmology of any scale-invariance modified gravity theory should be carefully checked with various tests. In the current work, two toy cyclic cosmological models with negative time-varying cosmological constant and no future singularity has been presented in the context of Conformal Bohm-de Broglie gravity.\par

\quad So many formulas for the positive time-dependent cosmological constant have been proposed in the literature. In this work, motivated by some recent observations, a formula for a negative time-dependent cosmological constant as a function of Hubble parameter has been suggested. A large class of purely phenomenological models with a generic time-varying cosmological constant exist in which the time dependence of $\Lambda(t)$ is introduced through ad hoc parametrization. Running vacuum models are theoretically motivated cosmological models where vacuum energy density evolves according to the renormalization group in quantum field theory in curved space-time. The time-dependent cosmological constant in the present work is not derived from the action but is introduced as a phenomenological description of dynamical vacuum energy, an approach that is widely adopted in cosmology.\par

\quad Physically acceptable results have been obtained only for the $\Lambda < 0$, while unphysical features with dominant negative energy densities through each cycle have been obtained for $\Lambda \geq 0$. Negative $\Lambda$ can still coexist with one evolving component of positive dark energy, although cosmic acceleration cannot be driven by it . The related impacts have been examined in light of cosmic microwave background, pre-DESI BAO dataset and recent JWST observations \cite{role5}.\par

\quad In the first model, a quintessence-dominated cosmic behavior has been revealed with no quintom behavior. Cosmic transit has been revealed by the sign flipping of deceleration parameter and cosmic pressure. The evolution of energy density shows that, besides its positivity for most of the evolution, it can also reach negative values in a small neighborhood of the bounce. For the matter bounce scenario, the energy density is positive all the time with quintom behavior near the bounce. Positive or vanishing cosmological constant leads to dominant negative energy density in both models. \par

\quad Both scalar potential and kinetic term are positive. Due to the difficulty of getting an expression for $V(\phi)$ in the present work, we have plotted $V(t)$ which becomes negative after the addition of quantum potential. In general, if $V(t)$ just becomes negative during evolution it is not enough to conclude AdS. Since the prediction of AdS spaces in string theory, negative potentials have attracted attention in cosmology and particle physics. It is generally accepted that a cyclic evolution will be easily obtained when a scalar field with a negative potential or a negative cosmological constant is added into a non-singular cosmological model, eventually leading to a recollapse.\par

\quad While most non-singular cosmological models require the NEC violation, it's highly preferable to avoid such violation if possible. No violation of NEC has been found for the first cyclic model in which only the SEC is violated. The NEC is partially violated in the second model only at the bounce. Because of the existence of quantum effects, we have considered the validity of the nonlinear energy conditions and found that they are all satisfied for the first model. The study in Ref.~\cite{kamel} where three different cosmological models have been constructed within conformal Bohm-de Broglie gravity considering $\Lambda > 0$ has been revisited and compared to the current negative $\Lambda$ study within same gravity.\par

\quad In the current work, the sign of the energy density has been treated as a first-level consistency requirement for selecting the negative $\Lambda$ solution. Nevertheless, in addition to this necessary physical condition, a comprehensive assessment should also address stability, perturbative behavior, and observational viability. While a full perturbation analysis has not been carried out, there are some non-trivial stability-related indicators such as: 1- the validity of the NEC. 2- the positivity of the kinetic term which indicates the absence of ghost-like behavior. So, beyond the requirement of a positive $\rho$, these two features suggest that the negative $\Lambda$ solution is not manifestly unstable at the background level, in contrast to many bouncing or cyclic scenarios that require strong violations of the NEC. A complete perturbation analysis would involve studying scalar/tensor modes, examining ghost and gradient stability, and investigating evolution through the bounce. While this lies beyond the scope of the present work, it represents a natural and necessary next step. While the negative $\Lambda$ solution is theoretically motivated (e.g., AdS vacuum, string landscape considerations), there exist recent phenomenological works and datasets exploring deviations from $\Lambda > 0$ (e.g., DESI-related analysis \cite{role5}). In the current work, the effective behavior mimics quintessence with dynamical equation of state, and the cyclic evolution avoids eternal acceleration which is itself a known conceptual issue in $\Lambda$CDM. So, while negative $\Lambda$ is not a confirmed observational result, it is a viable phenomenological solution worth exploring.

\quad Finally, we describe the future direction and outlooks of the current study. While the present work has focused on a toy cyclic cosmological model within scale-invariance modified gravity, several directions remain open for further investigation. A natural extension is to confront the model with a broader range of observational constraints, including updated BAO measurements, CMB polarization data, and forthcoming high-redshift probes. High redshift observations from the JWST and future 21-cm experiments may further probe the early cycle dynamics and potential deviations from the standard $\Lambda$CDM evolution. A systematic comparison with these data-sets will be essential to determine whether cyclic behavior driven by a negative time-dependent cosmological constant can be distinguished observationally from conventional dark-energy models. It would also be important to explore the stability of the cyclic solutions against perturbations and to examine the behavior of cosmological perturbations across the AdS–dS transitions. Finally, extending the model beyond the toy level to more realistic matter contents and interactions could clarify whether scale-invariant modified gravity can offer a consistent and predictive framework for non-singular cyclic cosmology.


\section*{Acknowledgements} 
The work of KB was supported in part by the JSPS KAKENHI Grant Numbers 24KF0100, 25KF0176, and Competitive Research Funds for Fukushima University Faculty (25RK011). 



\end{document}